\documentclass[11pt,a4paper,fleqn]{article}

\usepackage{lscape,exscale,amsthm,enumerate}
\usepackage[intlimits]{amsmath}
\usepackage{rawfonts}
\usepackage{latexsym}
\usepackage[cp850]{inputenc}
\usepackage{epsfig}
\usepackage{bbm}

\RequirePackage[OT1]{fontenc}
\RequirePackage{amsthm,amsmath}
\RequirePackage[numbers]{natbib}
\RequirePackage[colorlinks,citecolor=blue,urlcolor=blue]{hyperref}
\usepackage{stackrel}

\makeatletter
\newsavebox\myboxA
\newsavebox\myboxB
\newlength\mylenA

\newcommand*\xoverline[2][0.75]{%
    \sbox{\myboxA}{$\m@th#2$}%
    \setbox\myboxB\null
    \ht\myboxB=\ht\myboxA%
    \dp\myboxB=\dp\myboxA%
    \wd\myboxB=#1\wd\myboxA
    \sbox\myboxB{$\m@th\overline{\copy\myboxB}$}
    \setlength\mylenA{\the\wd\myboxA}
    \addtolength\mylenA{-\the\wd\myboxB}%
    \ifdim\wd\myboxB<\wd\myboxA%
       \rlap{\hskip 0.5\mylenA\usebox\myboxB}{\usebox\myboxA}%
    \else
        \hskip -0.5\mylenA\rlap{\usebox\myboxA}{\hskip 0.5\mylenA\usebox\myboxB}%
    \fi}
\makeatother

\numberwithin{equation}{section}
\theoremstyle{plain}
\newtheorem{theorem}{Theorem}[section]

\newtheorem{proposition}{Proposition}[section]
\newtheorem{lemma}{Lemma}[section]
\newtheorem{corollary}{Corollary}[section]

\newcommand{\bol}[1]{\mbox{\boldmath$#1$}}

\newcommand{\bSigma}{\bol{\Sigma}}

\newcommand{\bm}{\bol{\mu}}

\newcommand{\bx}{\mathbf{X}}

\newcommand{\by}{\mathbf{Y}}

\newcommand{\bC}{\mathbf{C}}

\newcommand{\bR}{\mathbf{R}}

\newcommand{\bw}{\mathbf{w}}

\newcommand{\bi}{\mathbf{1}}

\newcommand{\bI}{\mathbf{I}}

\newcommand{\bA}{\bol{A}}

\newcommand{\bS}{\mathbf{S}}

\usepackage{float}

\begin{document}

\begin{center}
\vspace*{2cm} \noindent {\bf \large Estimation of the Global Minimum Variance Portfolio in High Dimensions}\\
\vspace{1cm} \noindent {\sc  Taras Bodnar$^{a}$, Nestor Parolya$^{b}$ and Wolfgang Schmid$^{c}$
\footnote{Corresponding author. E-mail address: schmid@europa-uni.de} }\\
\vspace{1cm} {\it \footnotesize $^a$ Department of Mathematics, Stockholm University,  Roslagsv\"{a}gen 101, SE-10691 Stockholm, Sweden}\\
{\it \footnotesize $^b$ Department of Empirical Economics (Econometrics), Leibniz University Hannover, D-30167 Hannover, Germany}  \\
{\it \footnotesize $^c$ Department of Statistics, European University Viadrina, PO Box 1786, 15207 Frankfurt (Oder), Germany}
\end{center}

\begin{abstract}
We estimate the global minimum variance (GMV) portfolio in the high-dimensional case using results from random matrix theory. This approach leads to a shrinkage-type estimator which is distribution-free and it is optimal in the sense of minimizing the out-of-sample variance. Its asymptotic properties are investigated assuming that the number of assets $p$ depends on the sample size $n$ such that $\frac{p}{n}\rightarrow c\in (0,+\infty)$ as $n$ tends to infinity. The results are obtained under weak assumptions imposed on the distribution of the asset returns, namely it is only required the fourth moments existence. Furthermore, we make no assumption on the upper bound of the spectrum of the covariance matrix. As a result, the theoretical findings are also valid if the dependencies between the asset returns are described by a factor model which appears to be very popular in financial literature nowadays. This is also well-documented in a numerical study where the small- and large-sample behavior of the derived estimator are compared with existing estimators of the GMV portfolio. The resulting estimator shows significant improvements and it turns out to be robust to the deviations from normality.
\end{abstract}


\section{Introduction}

Since Markowitz (1952) presented his seminal work about portfolio selection, this topic has become a very fast growing branch of finance. One of Markowitz's ideas was the minimization of the portfolio variance subject to the budget constraint. This approach leads to the well-known and frequently used portfolio, the global minimum variance portfolio (GMV). There is a great amount of papers dealing with the GMV portfolio (see, e.g., Jagannathan and Ma (2003), Ledoit and Wolf (2003), Okhrin and Schmid (2006), Kempf and Memmel (2006), Bodnar and Schmid (2008), Frahm and Memmel (2010) among others). We remind that the GMV portfolio is the unique solution of the following optimization problem
\begin{equation}\label{VMIN}
\bw^{\prime}\bSigma_n\bw\rightarrow min~~\text{subject to}~~\bw^\prime\bi=1\,,
\end{equation}
where $\bw=(w_1,\ldots,w_p)^\prime$ denotes the vector of portfolio weights, $\bi$ is a suitable vector of ones, and $\bSigma_n$ stands for the covariance matrix of the asset returns. Note that in our paper $p$ is a function of the sample size $n$ and thus the covariance matrix depends on $n$ as well. This is shown by the index $n$.
The solution of (\ref{VMIN}) is given by
\begin{equation}\label{GMV}
\bw_{GMV}=\dfrac{\bSigma_n^{-1}\bi}{\bi^\prime\bSigma_n^{-1}\bi}.
\end{equation}
The GMV portfolio (\ref{GMV}) has the smallest variance over all portfolios. It is also used in multi-period portfolio choice problems (see, e.g., Brandt (2010)). Although this portfolio possesses several nice theoretical properties, some problems arise when the uncertainty about the parameters of the asset return distribution is taken into account. Indeed, we do not know the population covariance matrix in practice and, thus, it has to be suitably estimated. Consequently, the estimation of the GMV portfolio is strongly connected with the estimation of the covariance matrix of the asset returns.

The traditional estimator is a commonly used possibility for the estimation of the GMV portfolio (\ref{GMV}). This traditional estimator is constructed by replacing in (\ref{GMV}) the covariance matrix $\bSigma_n$ by its sample counterpart $\bS_n$. Okhrin and Schmid (2006) derived the distribution of the traditional estimator and studied its properties under the assumption that the asset returns follow a multivariate normal distribution, whereas Kempf and Memmel (2006) analyzed its conditional distributional properties. Furthermore, Bodnar and Schmid (2009) derived the distribution of the main characteristics of the sample GMV portfolio, namely its variance and its expected return.

The traditional estimator is not a bad choice if the number of assets $p$ is fixed and it is significantly smaller than the number of observations $n$ in the sample. This case is often used in statistics and it is called standard asymptotics (see, Le Cam and Yang (2000)). In that case the traditional estimator is a consistent estimator for the GMV portfolio and it is asymptotically normally distributed (Okhrin and Schmid (2006)). As a result, for a small fixed dimension  $p\in\{5,10,15\}$ we can use the sample estimator but it is not fully clear what to do if the number of assets in the portfolio is extremely large, say $p\in\{100, 500, 1000\}$, comparable to $n$. Here we are in the situation when both the number of assets $p$ and the sample size $n$ tend to infinity. This double asymptotics has an interpretation when $p$ and $n$ are of comparable size. More precisely, when $p/n$ tends to a concentration ratio $c>0$. This type of asymptotics is known as high-dimensional asymptotics or "Kolmogorov" asymptotics (see, Bai and Silverstein (2010)). Under the high-dimensional asymptotics the traditional estimator behaves very unpredictable and it is far from the optimum one. It tends to underestimate the risk (see, El Karoui (2010), Bai and Shi (2011)). In general, the traditional estimator is worse for larger values of the concentration ratio $c$. Imposing the assumption of a factor structure on the asset returns this problem was resolved in an efficient way by Bai (2003), Fan et al. (2008), Fan et al. (2012), Fan et al. (2013), etc. Nevertheless, if the factor structure is not present the question of high-dimensionality remains open.

Further estimators for the weights of the GMV portfolio have been proposed in this situation. DeMiguel et al. (2009) suggested to involve some additional portfolio constraints in order to avoid the curse of dimensionality. On the other hand, shrinkage estimators can be used which are biased but can significantly reduce the risk of the portfolio by minimizing its mean-square error. The general shrinkage estimator is a convex combination of the traditional estimator and a known target (for the GMV portfolio it can be the naive equally weighted portfolio). They were first considered by Stein (1956). Recently, various authors showed that shrinkage estimators for the portfolio weights indeed lead to better results (see, e.g., Golosnoy and Okhrin (2007), Frahm and Memmel (2010)). In particular, Golosnoy and Okhrin (2007) considered a multivariate shrinkage estimator by shrinking the portfolio weights themselves but not the whole sample covariance matrix. The same idea was used by Frahm and Memmel (2010) who constructed a feasible shrinkage estimator for the GMV portfolio which dominates the traditional one.  
There are several problems with these estimators: first, the normal distribution is usually imposed; second, dominating does not mean optimal; and third, the large dimensional behavior (large $p$ and large $n$) seems not to be acceptable. 

The aim of the paper is to derive a feasible and simple estimator for the GMV portfolio which is optimal, in some sense, for small and large sample sizes and which is distribution-free as well. For that purpose  we construct an optimal shrinkage estimator, study its asymptotic properties and estimate unknown quantities consistently. The estimator is obtained using random matrix theory, a fast growing branch of probability theory. The main result of this theory was proved by Mar$\breve{\text{c}}$enko and Pastur (1967) and further extended under very general conditions by Silverstein (1995). Nowadays it is called Mar$\breve{\text{c}}$enko-Pastur equation. Its importance arises in many areas of science because it shows how the real covariance matrix and its sample estimate are connected with each other. Knowing this information we can build suitable estimators for high-dimensional quantities.

The rest of the paper is organized as follows. In Section 2 we present a shrinkage estimator for the GMV portfolio which is optimal in terms of minimizing the out-of-sample variance. The asymptotic behavior of the resulting shrinkage intensity is investigated for $c<1$ in Section 2.1 and in case of $c>1$ in Section 2.2 where it is shown that the shrinkage intensity tends almost surely to a deterministic quantity when both the sample size and the portfolio dimension increase. This result allows us to determine an oracle estimator of the GMV portfolio, while the corresponding bona fide estimator is presented in Section 2.3. In Section 3 we provide a simulation study for different values of $c\in(0, +\infty)$ and under various distributional assumptions imposed on the data generating process. Here, the performance and the convergence rate of the derived shrinkage estimator are compared with existing estimators of the GMV portfolio. The results of our empirical study are given in Section 4 where we apply the suggested estimator as well as the existing estimators to real data consisting of returns on assets included in the S\&P 500 (Standard \& Poor's 500) index. Section 5 summarizes all of the obtained results. The lengthy proofs are moved to the appendix (Section 6).

\section{Optimal shrinkage estimator for the GMV portfolio}

Let $\by_n=(\mathbf{y}_1,\mathbf{y}_2,...,\mathbf{y}_n)$ be the $p \times n$ data matrix which consists of $n$ vectors of the returns on $p\equiv p(n)$ assets. Let $E(\mathbf{y}_i)=\bm_n$ and $Cov(\mathbf{y}_i)=\bSigma_n$ for $i \in 1,...,n$. We assume that $p/n\rightarrow c\in (0, +\infty)$ as $n\rightarrow\infty$. This type of limiting behavior is also denoted as "large dimensional asymptotics" or "the Kolmogorov asymptotics". In this case the traditional estimators perform poor or even very poor and tend to over/underestimate the unknown parameters of the asset returns, i.e., the mean vector and the covariance matrix.

Throughout the paper it is assumed that it exists a $p\times n$ random matrix $\bx_n$ which consists of independent and identically distributed (i.i.d.) real random variables with zero mean and unit variance such that
\begin{equation}\label{obs}
 \by_n= \bm_n \bi^\prime + \bSigma_n^{\frac{1}{2}}\bx_n \,.
 \end{equation}
It is noted that the observation matrix $\by_n$ consists of dependent rows although its columns are independent. The assumption of the independence of the columns can further be weakened by controlling the growth of the number of dependent entries, while no specific distributional assumptions are imposed on the elements of $\by_n$ (see, Friesen et al.(2013)).

The two main assumptions which are used throughout the paper are
\begin{enumerate}
\item[(A1)] The covariance matrix of the asset returns $\bSigma_n$ is a nonrandom $p$-dimensional positive definite matrix.

\item[(A2)] The elements of the matrix $\bx_n$ have uniformly bounded $4+\varepsilon$ moments for some $\varepsilon>0$.
\end{enumerate}

These two regularity conditions are very general and they fit many practical situations. The assumption (A1) is common for financial and statistical problems. It does not impose a strong restriction on the data-generating process, whereas the assumption (A2) is purely technical. Moreover, it seems to influence only the convergence rate of the proposed estimator (see, e.g. Rubio et al. (2012)).\\
The sample covariance matrix is given by
\begin{equation}\label{samplecov}
 \bS_n=\dfrac{1}{n}\by_n(\bI-\frac{1}{n}\bi\bi^\prime)\by_n^{\prime}=\dfrac{1}{n}\bSigma_n^{\frac{1}{2}}\bx_n(\bI-\frac{1}{n}\bi\bi^\prime)\bx_n^{\prime}\bSigma_n^{\frac{1}{2}}\,,
 \end{equation}
where the symbol $\bI$ stands for the identity matrix of an appropriate dimension.

\subsection{Oracle estimator. Case $c<1$}
The traditional estimator for the GMV portfolio is obtained by replacing the unknown population covariance matrix $\bSigma_n$ in \eqref{GMV} by the estimator \eqref{samplecov} . This leads to
\begin{equation}\label{GMV_sam_est}
\hat{\bw}_{GMV}=\dfrac{\bS_n^{-1}\bi}{\bi^\prime\bS_n^{-1}\bi}.
\end{equation}

Next, we derive the optimal shrinkage estimator for the GMV portfolio weights by optimizing with respect to the shrinkage parameter $\alpha_n$ and fixing some target portfolio $\mathbf{b}_n$. Its distributional properties are studied after that. The general shrinkage estimator (GSE) for $c\in (0, 1)$ is defined by
\begin{equation}\label{gse}
\hat{\bw}_{GSE}=\alpha_n\dfrac{\bS_n^{-1}\bi}{\bi^\prime\bS_n^{-1}\bi}+(1-\alpha_n)\mathbf{b}_n~~\text{with}~\mathbf{b}_n^\prime\bi=1\,
\end{equation}
where $\mathbf{b}_n\in\mathbbm{R}^p$ is a given nonrandom (or random but independent of the actual observation vector $\mathbf{y}_n$, i.e., the last column of $\by_n$) vector. 
No assumption is imposed on the shrinkage intensity $\alpha_n$ which is the object of our interest. The aim is to find the optimal shrinkage intensity $\alpha_n$. 
for a given target portfolio $\mathbf{b}_n$ which minimizes the out-of-sample risk\\

\vspace{-10mm}

\begin{eqnarray}\label{risk}
L&=&||\bSigma^{\frac{1}{2}}_n(\hat{\bw}_{GSE}(\alpha_n)-\bw_{GMV})||=(\hat{\bw}_{GSE}(\alpha_n)-\bw_{GMV})^\prime\bSigma_n(\hat{\bw}_{GSE}(\alpha_n)-\bw_{GMV})\,,
\end{eqnarray}
(see, e.g., Frahm and Memmel (2010), Rubio et al. (2012)). The loss function (\ref{risk}) can be rewritten as
\begin{eqnarray}\label{minvar}
&&L=\hat{\bw}^\prime_{GSE}(\alpha_n)\bSigma_n\hat{\bw}_{GSE}(\alpha_n)-\sigma_{GMV}^2\,,
\end{eqnarray}
where $\sigma_{GMV}^2=\dfrac{1}{\bi^\prime\bSigma_n^{-1}\bi}$ is the population variance of the GMV portfolio and $\hat{\bw}^\prime_{GSE}(\alpha_n)\bSigma_n\hat{\bw}_{GSE}(\alpha_n)$ is known as the out-of-sample variance of the portfolio with the weights $\hat{\bw}_{GSE}(\alpha_n)$.

 Using (\ref{gse}) we want to solve the following optimization problem
\begin{eqnarray}\label{optm}
\min_{\alpha_n} L= \min_{\alpha_n} \alpha_n^2\sigma^2_{S}+2\alpha_n(1-\alpha_n)\dfrac{1}{\bi^\prime\bS_n^{-1}\bi}\bi^\prime\bS_n^{-1}\bSigma_n\mathbf{b}_n
+(1-\alpha_n)^2\mathbf{b}_n^{\prime}\bSigma_n\mathbf{b}_n-\sigma_{GMV}^2\,,
\end{eqnarray}
where
\begin{align}\label{xyz}
&\sigma^2_{S}=\dfrac{\bi^\prime\bS_n^{-1}\bSigma_n\bS_n^{-1}\bi}{(\bi^\prime\bS_n^{-1}\bi)^2}\,
\end{align}
is the out-of-sample variance of the traditional estimator for the GMV portfolio weights. Taking the derivative of $L$ with respect to $\alpha_n$ and setting it equal to zero we get
\begin{equation}\label{der2}
\dfrac{\partial L}{\partial\alpha_n}=\alpha_n\sigma^2_{S}+(1-2\alpha_n)\dfrac{\bi^\prime\bS_n^{-1}\bSigma_n\mathbf{b}_n}{\bi^\prime\bS_n^{-1}\bi}-(1-\alpha_n)\mathbf{b}_n^{\prime}\bSigma_n\mathbf{b}_n=0\,.
\end{equation}
 From the last equation it is easy to find the optimal shrinkage intensity $\alpha_n^*$ given by
 \begin{equation}\label{alfa} \alpha_n^*=\dfrac{\mathbf{b}_n^{\prime}\bSigma_n\mathbf{b}_n-\dfrac{\bi^\prime\bS_n^{-1}\bSigma_n\mathbf{b}_n}{\bi^\prime\bS_n^{-1}\bi}}{\sigma^2_{S}
 -2\dfrac{\bi^\prime\bS_n^{-1}\bSigma_n\mathbf{b}_n}{\bi^\prime\bS_n^{-1}\bi}+\mathbf{b}_n^{\prime}\bSigma_n\mathbf{b}_n}
 =\dfrac{\left(\mathbf{b}_n-\dfrac{\bS^{-1}_n\bi}{\bi^\prime\bS^{-1}_n\bi}\right)^\prime\bSigma_n\mathbf{b}_n}
 {\left(\mathbf{b}_n-\dfrac{\bS^{-1}_n\bi}{\bi^\prime\bS^{-1}_n\bi}\right)^\prime\bSigma_n\left(\mathbf{b}_n-\dfrac{\bS^{-1}_n\bi}{\bi^\prime\bS^{-1}_n\bi}\right)}\,.
 \end{equation}

In order to ensure that $\alpha_n^*$ is the minimizer of (\ref{optm}) we calculate the second derivative of $L$ which has to be positive. It holds that
\begin{equation}\label{secder}
\dfrac{\partial^2 L}{\partial\alpha^2_n}=\sigma^2_{S}-2\bi^\prime\bS_n^{-1}\bSigma_n\mathbf{b}_n\dfrac{1}{\bi^\prime\bS_n^{-1}\bi}+\mathbf{b}_n^{\prime}\bSigma_n\mathbf{b}_n
=\left(\mathbf{b}_n-\dfrac{\bS^{-1}_n\bi}{\bi^\prime\bS^{-1}_n\bi}\right)^\prime\bSigma_n\left(\mathbf{b}_n-\dfrac{\bS^{-1}_n\bi}{\bi^\prime\bS^{-1}_n\bi}\right)>0\,
\end{equation}
 almost surely. The last inequality is always true because of the positive definiteness of the matrix $\bSigma_n$ and the fact that $\mathbf{b}_n=\dfrac{\bS^{-1}_n\bi}{\bi^\prime\bS^{-1}_n\bi}$ with probability zero.

In Theorem \ref{th1} we show that the optimal shrinkage intensity $\alpha^*_n$ is almost surely asymptotically equivalent to a nonrandom quantity $\alpha^* \in [0, 1]$ under the large-dimensional asymptotics $\dfrac{p}{n}\rightarrow c\in(0, 1)$. Let $\sigma_{\mathbf{b}_n}=\mathbf{b}_n^{\prime}\bSigma_n\mathbf{b}_n $ be the variance of the target portfolio and let
\[R_{\mathbf{b}_n}=\dfrac{\sigma_{\mathbf{b}_n}^2-\sigma_{GMV}^2}{\sigma_{GMV}^{2}}\]
be the relative loss of the target portfolio $\mathbf{b}_n$.

\vspace{0.5cm}
\begin{theorem}\label{th1}
Assume (A1)-(A2). Let $0< M_l \le \sigma_{GMV}^2 \le \sigma_{\mathbf{b}_n}^2 \le M_u <\infty $ for all $n$. Then it holds that

\vspace{-6mm}

\begin{equation}\label{ainfty}
\alpha^*_n\overset{a.s.}{\longrightarrow}\alpha^*=\dfrac{(1-c)R_{\mathbf{b}}}{c+(1-c)R_{\mathbf{b}}}~~\text{for} ~~ \dfrac{p}{n}\rightarrow c\in(0, 1)~~ \text{as} ~~ n\rightarrow\infty\,,
\end{equation}
where $R_{\mathbf{b}}$ is the limit of $R_{\mathbf{b}_n}$.
Additionally, the out-of-sample variance $\sigma^2_S$ of the traditional estimator for the GMV portfolio possesses the following asymptotic behavior
\begin{equation}\label{outsampvariance}
\sigma^2_S\overset{a.s.}{\longrightarrow}\dfrac{1}{1-c}\sigma_{GMV}^2 ~~\text{for} ~~ \dfrac{p}{n}\rightarrow c\in(0, 1)~~ \text{as} ~~ n\rightarrow\infty\,.
\end{equation}
\end{theorem}

\vspace{0.5cm}
The proof of Theorem \ref{th1} is given in the Appendix. Theorem \ref{th1} provides us important information about the optimal shrinkage estimator of the GMV portfolio. Especially, the application of Theorem \ref{th1} immediately leads to consistent estimators for $\alpha^*_n$, $\sigma^2_{GMV}$, and $\sigma^2_{S}$ which are presented in Section 2.3 below. It is remarkable to note that the assumption $0< M_l\le \sigma_{GMV}^2 \le \sigma_{\mathbf{b}_n}^2 \le M_u <\infty $ is natural for financial markets. It ensures that the population variance of the GMV portfolio has a lower bound which is in-line with the Capital Asset Pricing Model since the portfolio variance cannot be smaller than the market risk (see, e.g., Elton et al. (2007, Chapter 7)). Moreover, the assumption of the boundedness of the variance of the target portfolio $\sigma_{\mathbf{b}_n}^2$ is also well acceptable because it makes no sense to shrink to a portfolio with infinite variance. Most importantly, this condition also holds even if the largest eigenvalue of the covariance matrix is unbounded. Such a situation is present if the asset returns follow a factor model which is a very popular approach in financial literature nowadays (see, e.g., Fan et al. (2008), Fan et al. (2012)). It is worth pointing out that the same result is true if we assume instead of $0< M_l\le \sigma_{GMV}^2 \le \sigma_{\mathbf{b}_n}^2 \le M_u <\infty$ the boundedness of the spectral norm of the population covariance matrix, i.e., the uniformly bounded maximum eigenvalue of $\bSigma_n$.

The answer on the question about the performance of the traditional and the optimal shrinkage estimator for the GMV portfolio is given in Corollary \ref{cor1}.

\vspace{0.5cm}
\begin{corollary}\label{cor1}
\begin{enumerate}[(a)]
\item
 Under the assumptions of Theorem \ref{th1}, we get for the relative loss of the traditional estimator for the GMV portfolio
\begin{equation}\label{outsamp1}
R_S=\dfrac{\sigma^2_{S}-\sigma_{GMV}^2}{\sigma_{GMV}^2}\overset{a.s.}{\longrightarrow}\dfrac{c}{1-c}~~\text{for} ~~ \dfrac{p}{n}\rightarrow c\in(0, 1)~~ \text{as} ~~ n\rightarrow\infty\,.
\end{equation}
\item
Under the assumptions of Theorem \ref{th1}, we get for the relative loss of the optimal shrinkage estimator for the GMV portfolio
\begin{equation}\label{outsamp2}
R_{GSE}=\dfrac{\hat{\bw}_{GSE}^T\bSigma_n\hat{\bw}_{GSE}-\sigma_{GMV}^2}{\sigma_{GMV}^2}\overset{a.s.}{\longrightarrow}~~
(\alpha^*)^2\dfrac{c}{1-c}+(1-\alpha^*)^2R_{\mathbf{b}}~~\text{for} ~~ \dfrac{p}{n}\rightarrow c\in(0, 1)~~ \text{as} ~~ n\rightarrow\infty\,.
\end{equation}
\end{enumerate}
\end{corollary}


Corollary \ref{cor1} is a straightforward consequence of Theorem \ref{th1}. Moreover, its first part generalizes the result of Frahm and Memmel (2010, Theorem 7) to an arbitrary distribution of the asset returns. Using Corollary \ref{cor1}(a) we can plot the behavior of the relative loss of the traditional estimator for the GMV portfolio as a function of the concentration ratio $c$ only, while the relative loss of the optimal shrinkage portfolio additionally depends on the relative loss of the target portfolio. Furthermore, from both parts of Corollary \ref{cor1} we get
\[R_{GSE}\overset{a.s.}{\longrightarrow} (\alpha^*)^2R_S+(1-\alpha^*)^2R_{\mathbf{b}}~~\text{for} ~~ \dfrac{p}{n}\rightarrow c\in(0, 1)~~ \text{as} ~~ n\rightarrow\infty\,,
\]
i.e., the relative loss of the optimal shrinkage estimator for the GMV portfolio can asymptotically be presented as a linear combination of the relative loss of the traditional estimator and the relative loss of the target portfolio. Because $\alpha^* \rightarrow 0$ as $c \rightarrow 1_{-}$\footnote{Further in paper, $c\rightarrow1_{-}$ and $c\rightarrow1_{+}$ denote the left and right limits to the point $1$, respectively.} and
\[
(\alpha^*)^2\dfrac{c}{1-c}=\frac{(1-c)cR_{\mathbf{b}}^2}{(c+(1-c)R_{\mathbf{b}})^2} \rightarrow 0 ~~ \text{as} ~~ c \rightarrow 1_{-}\,,
\]
we get that $R_{GSE} \rightarrow R_{\mathbf{b}}\leq\frac{M_u-M_l}{M_l}$ as $c \rightarrow 1_{-}$, whereas the relative loss of the traditional estimator tends to infinity.

%

\begin{figure}[H]
\begin{center}
\includegraphics[scale=0.359]{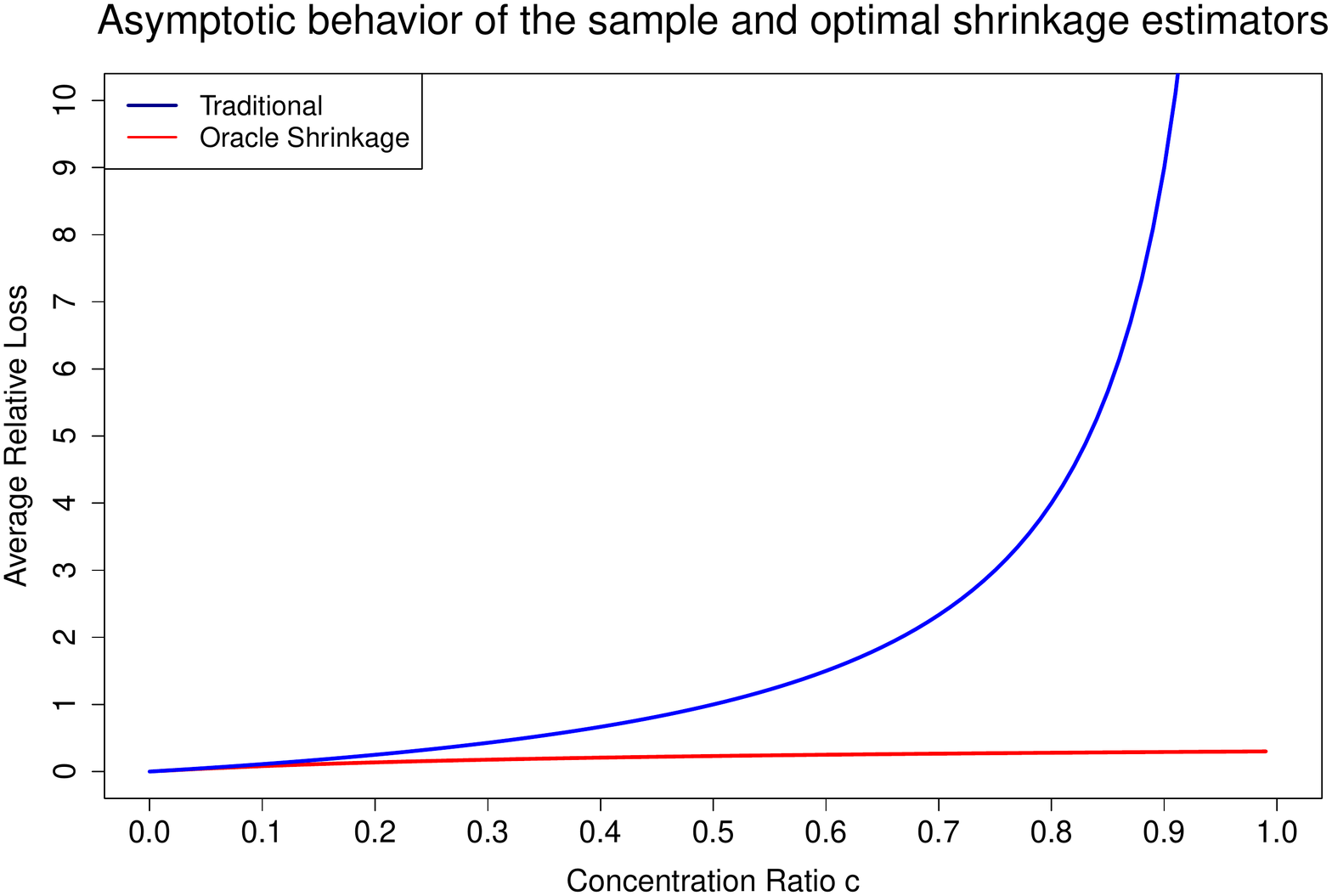}
\vspace{-5mm}
\caption{Asymptotic relative loss of the traditional and the oracle shrinkage estimators as a function of the concentration ratio $p/n=c<1$. }
\label{Fig1}
\end{center}
\end{figure}

Figure \ref{Fig1} presents the behavior of the traditional and the proposed oracle estimators of the GMV portfolio weights for different values of $c\in(0,1)$. The covariance matrix $\bSigma_n$ is taken as a $200\times200$-dimensional matrix where we have taken $20\%$ of the eigenvalues equal to $3$, $40\%$ equal to $1$, and $40\%$ equal to $0.5$. The matrix of eigenvectors $\mathbf{V}=(\mathbf{v}_1,\ldots,\mathbf{v}_p)^\prime$ is generated from the Haar distribution\footnote{If $\mathbf{V}$ has a Haar measure over the orthogonal matrices, then for any unit vector $\mathbf{x}\in\mathbbm{R}^p$, $\mathbf{V}\mathbf{x}$ has a uniform distribution over the unit sphere $S_p=\left\{\mathbf{x}\in\mathbbm{R}^p ; ||\mathbf{x}||=1\right\}$.} 
The target portfolio is chosen as the equally weighted portfolio, i.e. $\mathbf{b}_n=1/p\bi$. In the figure we observe that the asymptotic relative loss of the traditional estimator for the GMV portfolio has a singularity point at one. The loss of the traditional estimator is relatively small up to $c=0.2$ but thereafter, as $p/n\rightarrow 1$, it rises hyperbolically to infinity. In contrast to the traditional estimator of the GMV portfolio weights, the suggested optimal shrinkage estimator has a constant asymptotic relative loss which is always smaller than $0.5$. This result is in-line with the theoretical findings discussed around Corollary \ref{cor1}.

\begin{figure}[H]
\begin{center}
\includegraphics[scale=0.4]{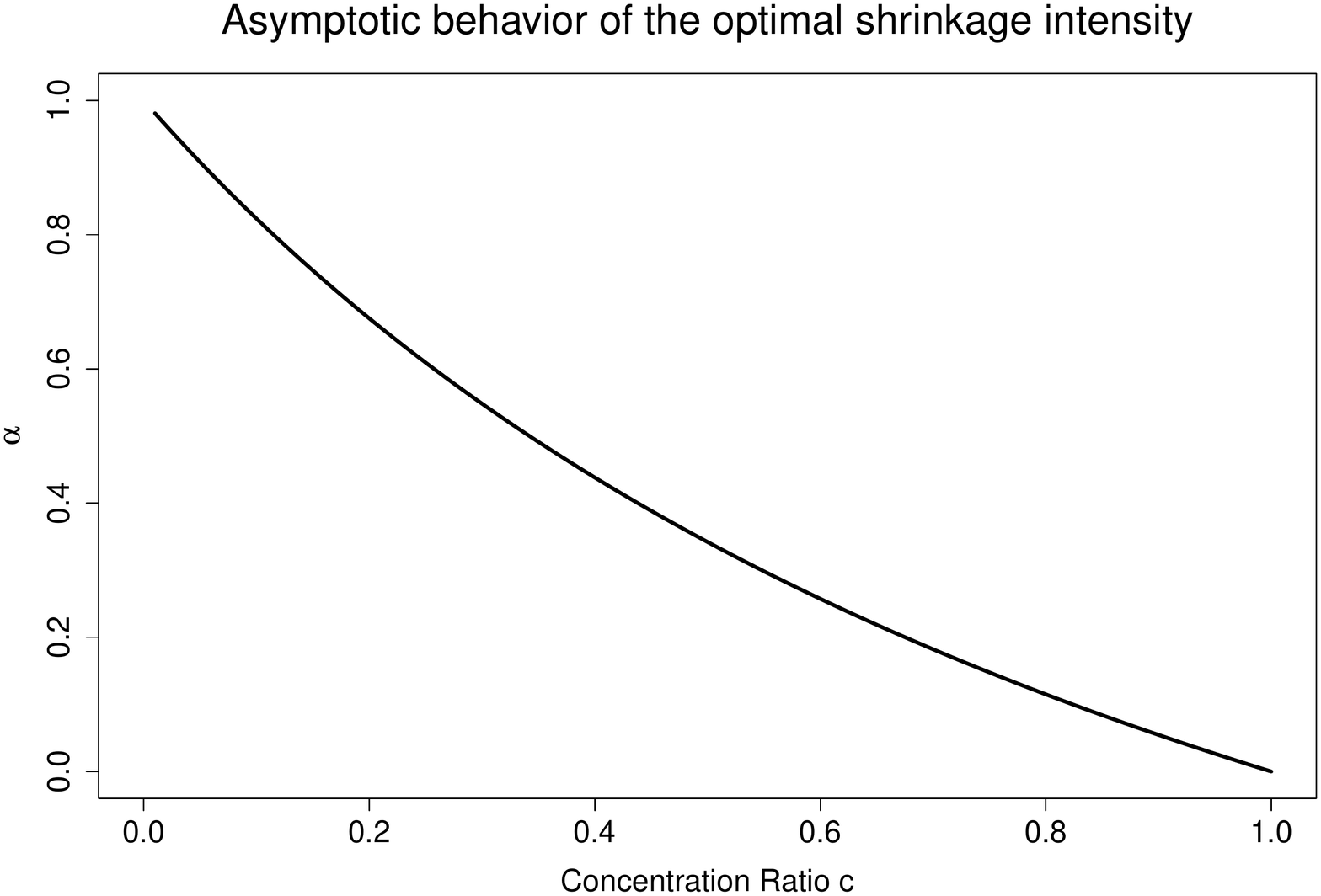}
\vspace{-5mm}
\caption{Asymptotic behavior of the optimal shrinkage intensity $\alpha^*$ as a function of the concentration ratio $p/n=c<1$.}
\label{Fig2}
\end{center}
\end{figure}

\begin{center} Figures \ref{Fig1} and \ref{Fig2} above here \end{center}

In Figure \ref{Fig2} we show the asymptotic behavior of the optimal shrinkage intensity $\alpha^*$ as a function of the concentration ratio $c\in(0, 1)$. The target portfolio $\mathbf{b}_n$ and the covariance matrix $\bSigma_n$ are the same as in Figure \ref{Fig1}. In the interval $c\in(0, 1)$ the optimal shrinkage intensity $\alpha^*$ is a nonlinearly decreasing (in a convex manner) function of the concentration ratio $c$. We observe that the optimal $\alpha^*$ tends to zero as $c$ approaches one and, thus, in the limiting case the only optimal choice would be the target portfolio $\mathbf{b}_n$.

\subsection{Oracle estimator. Case $c>1$}\label{secc1}
In case $c>1$, the sample covariance matrix $\bS_n$ is singular and its inverse does not exist anymore. Thus, we first have to find a reasonable replacement for $\bS_n^{-1}$. For the oracle estimator of the GMV portfolio weights we use the following generalized inverse of the sample covariance matrix $\bS_n$
\begin{equation}\label{geninverse}
\bS_n^*=\bSigma_n^{-1/2}(\bx_n\bx_n^\prime)^+\bSigma_n^{-1/2}\,,
\end{equation}
where $'+'$ denotes the Moore-Penrose inverse. It can be shown that $\bS_n^*$ is the generalized inverse satisfying $\bS_n^*\bS_n\bS_n^*=\bS_n^*$ and $\bS_n\bS_n^*\bS_n=\bS_n$.\footnote{Note that $\bS_n^*$ is not equal to the Moore-Penrose inverse because it does not satisfy the conditions $(\bS_n^*\bS_n)^\prime=\bS_n^*\bS_n$ and $(\bS_n\bS_n^*)^\prime=\bS_n\bS_n^*$. Nevertheless, in Section 2.3, where the bona fide estimator is constructed, we use the Moore-Penrose inverse of $\bS_n$ instead of $\bS_n^*$ in order to obtain a valuable approximation.} Obviously, in case $c<1$ the generalized inverse $\bS_n^*$ coincides with the usual inverse $\bS_n^{-1}$. Moreover, if $\bSigma_n$ is proportional to the identity matrix then $\bS_n^*$ coincides with the Moore-Penrose inverse $\bS_n^+$ calculated for $\bS_n$. It has also to be noted that $\bS_n^*$ cannot be determined in practice since it depends on the unknown matrix $\bSigma_n$. In this section, it is only used to determine an oracle estimator for the weights of the GMV portfolio, whereas the bona fide estimator is constructed in Section 2.3.

Based on $\bS_n^*$ in \eqref{geninverse}, the oracle traditional estimator for the GMV portfolio in case $c>1$ is first constructed and it is given by
\begin{equation}\label{trad+}
\hat{\bw}^*_{GMV}=\dfrac{\bS_n^{*}\bi}{\bi^\prime\bS_n^{*}\bi}\,.
\end{equation}

Next, we determine the oracle optimal shrinkage estimator for the GMV portfolio weights expressed as
\begin{equation}\label{gse+}
\hat{\bw}^*_{GSE}=\alpha^+_n\dfrac{\bS_n^{*}\bi}{\bi^\prime\bS_n^{*}\bi}+(1-\alpha^+_n)\mathbf{b}_n~~\text{with}~\mathbf{b}_n^\prime\bi=1\,.
\end{equation}
Similarly to Section 2.1, we deduce the optimal shrinkage intensity $\alpha^+_n$ given by
\begin{equation}\label{alfa+}
\alpha_n^+=\dfrac{\mathbf{b}_n^{\prime}\bSigma_n\mathbf{b}_n-\dfrac{\bi^\prime\bS_n^{*}\bSigma_n\mathbf{b}_n}{\bi^\prime\bS_n^{*}\bi}}{\sigma^2_{S^*}-2\dfrac{\bi^\prime\bS_n^{*}\bSigma_n\mathbf{b}_n}{\bi^\prime\bS_n^{*}\bi}+\mathbf{b}_n^{\prime}\bSigma_n\mathbf{b}_n}=\dfrac{\left(\mathbf{b}-\dfrac{\bS^{*}_n\bi}{\bi^\prime\bS^{*}_n\bi}\right)^\prime\bSigma_n\mathbf{b}}{\left(\mathbf{b}-\dfrac{\bS^{*}_n\bi}{\bi^\prime\bS^{*}_n\bi}\right)^\prime\bSigma_n\left(\mathbf{b}-\dfrac{\bS^{*}_n\bi}{\bi^\prime\bS^{*}_n\bi}\right)}\,,
 \end{equation}
where $\sigma^2_{S^*}=\bi^\prime\bS_n^*\bSigma_n\bS_n^*\bi/(\bi^\prime\bS_n^*\bi)^2$ is the oracle out-of-sample variance of the traditional estimator for the GMV portfolio. In Theorem \ref{th12} we present the asymptotic properties of the optimal $\alpha^+_n$ for $c>1$.

\begin{theorem}\label{th12}
Assume (A1)-(A2). Let $0< M_l \le \sigma_{GMV}^2 \le \sigma_{\mathbf{b}_n}^2 \le M_u <\infty $ for all $n$. Then it holds that

\vspace{-6mm}

\begin{equation}\label{ainfty+}
\alpha^+_n\overset{a.s.}{\longrightarrow}\alpha^+=\dfrac{(c-1)R_{\mathbf{b}}}{(c-1)^2+c+(c-1)R_{\mathbf{b}}}~~\text{for} ~~ \dfrac{p}{n}\rightarrow c\in(1, +\infty)~~ \text{as} ~~ n\rightarrow\infty\,,
\end{equation}
where $R_{\mathbf{b}}$ is the limit of $R_{\mathbf{b}_n}$.
Additionally, we get for the oracle out-of-sample variance $\sigma^2_{S^*}$ of the traditional estimator \eqref{trad+} for the GMV portfolio
\begin{equation}\label{outsampvariance+}
\sigma^2_{S^*}\overset{a.s.}{\longrightarrow}\dfrac{c^2}{c-1}\sigma_{GMV}^2 ~~\text{for} ~~ \dfrac{p}{n}\rightarrow c\in(1, +\infty)~~ \text{as} ~~ n\rightarrow\infty\,.
\end{equation}
\end{theorem}

The proof of Theorem \ref{th12} is given in the appendix. The asymptotic behavior of the relative loss calculated for the traditional oracle estimator of the GMV portfolio as well as for the oracle optimal shrinkage estimator is described in Corollary \ref{cor2}.
\vspace{0.5cm}
\begin{corollary}\label{cor2}
\begin{enumerate}[(a)]
\item
 Under the assumptions of Theorem \ref{th12}, we get for the relative loss of the oracle traditional estimator for the GMV portfolio
\begin{equation}\label{outsamp1+}
R_{S}^{*}=\dfrac{\sigma^2_{S^*}-\sigma_{GMV}^2}{\sigma_{GMV}^2}\overset{a.s.}{\longrightarrow}\dfrac{c^2-c+1}{c-1}~~\text{for} ~~ \dfrac{p}{n}\rightarrow c\in(1, +\infty)~~ \text{as} ~~ n\rightarrow\infty\,.
\end{equation}
\item Under the assumptions of Theorem \ref{th12}, we get for the relative loss of the oracle optimal shrinkage estimator for the GMV portfolio
\begin{equation}\label{outsamp2+}
R_{GSE}^*=\dfrac{(\hat{\bw}_{GSE}^*)^T\bSigma_n\hat{\bw}_{GSE}^*-\sigma_{GMV}^2}{\sigma_{GMV}^2}\overset{a.s.}{\longrightarrow}~~
(\alpha^+)^2R_{S}^{*}+(1-\alpha^+)^2R_{\mathbf{b}}~~\text{for} ~~ \dfrac{p}{n}\rightarrow c\in(0, 1)~~ \text{as} ~~ n\rightarrow\infty\,.
\end{equation}
\end{enumerate}
\end{corollary}

Similarly to the case $c<1$, the relative loss of the optimal shrinkage estimator for the GMV portfolio is a linear combination of the relative loss of the traditional estimator and the relative loss of the target portfolio. Furthermore, if $c \rightarrow 1_{+}$, the relative loss of the traditional estimator tends to infinite\footnote{The sample covariance matrix $\bS_n$ is ill-behaved and not invertible at the point $c=1$ because in that case its smallest eigenvalue is very near to zero.}, whereas for the relative loss of the shrinkage estimator we get
\[R_{GSE}^*\rightarrow \dfrac{(c-1)(c^2-c+1)R_{\mathbf{b}}^2}{((c-1)^2+c+(c-1)R_{\mathbf{b}})^2} +(1-\alpha^+)^2R_{\mathbf{b}} =R_{\mathbf{b}} ~~ \text{as} ~~ c \rightarrow 1_{+}\,,\]
which is bounded from above by $\frac{M_u-M_l}{M_l}$, i.e., it is finite.

\begin{figure}[H]
\begin{center}
\includegraphics[scale=0.4]{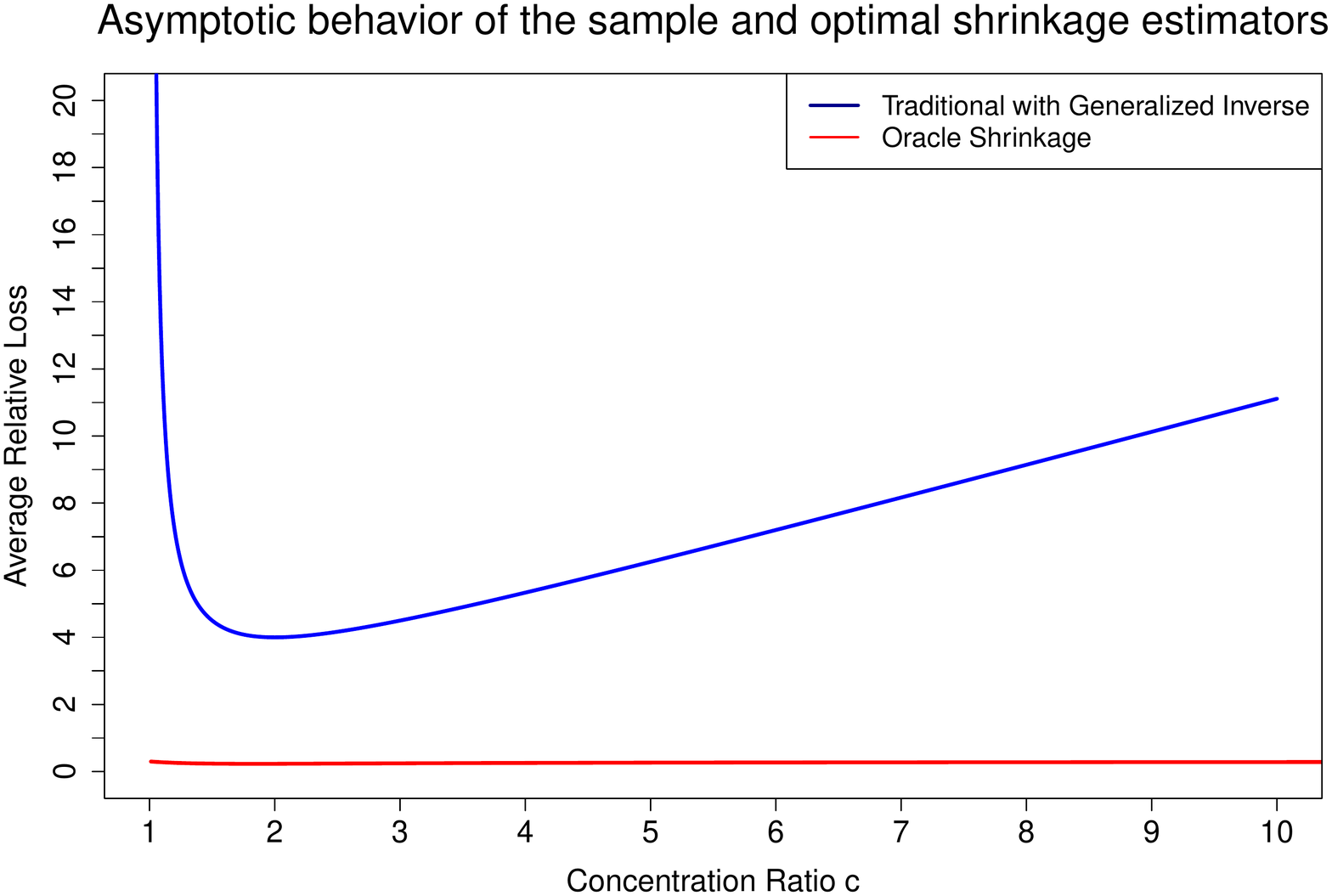}
\vspace{-5mm}
\caption{Asymptotic relative loss of the oracle traditional estimator with generalized inverse \eqref{gse+} and of the oracle optimal shrinkage estimator as a function of the concentration ratio $p/n=c>1$. }
\label{Fig3}
\end{center}
\end{figure}

Figure \ref{Fig3} presents the asymptotic performance of the oracle traditional estimator and of the proposed oracle optimal shrinkage estimator for the GMV portfolio in case $c>1$. A considerable improvement is present when the oracle optimal shrinkage estimator is applied where the average loss is always smaller than $1$. In contrast, the average loss of the oracle traditional estimator possesses always larger values with a minimum of about $4$ reached around $c=2$.

\begin{figure}[H]
\begin{center}
\includegraphics[scale=0.4]{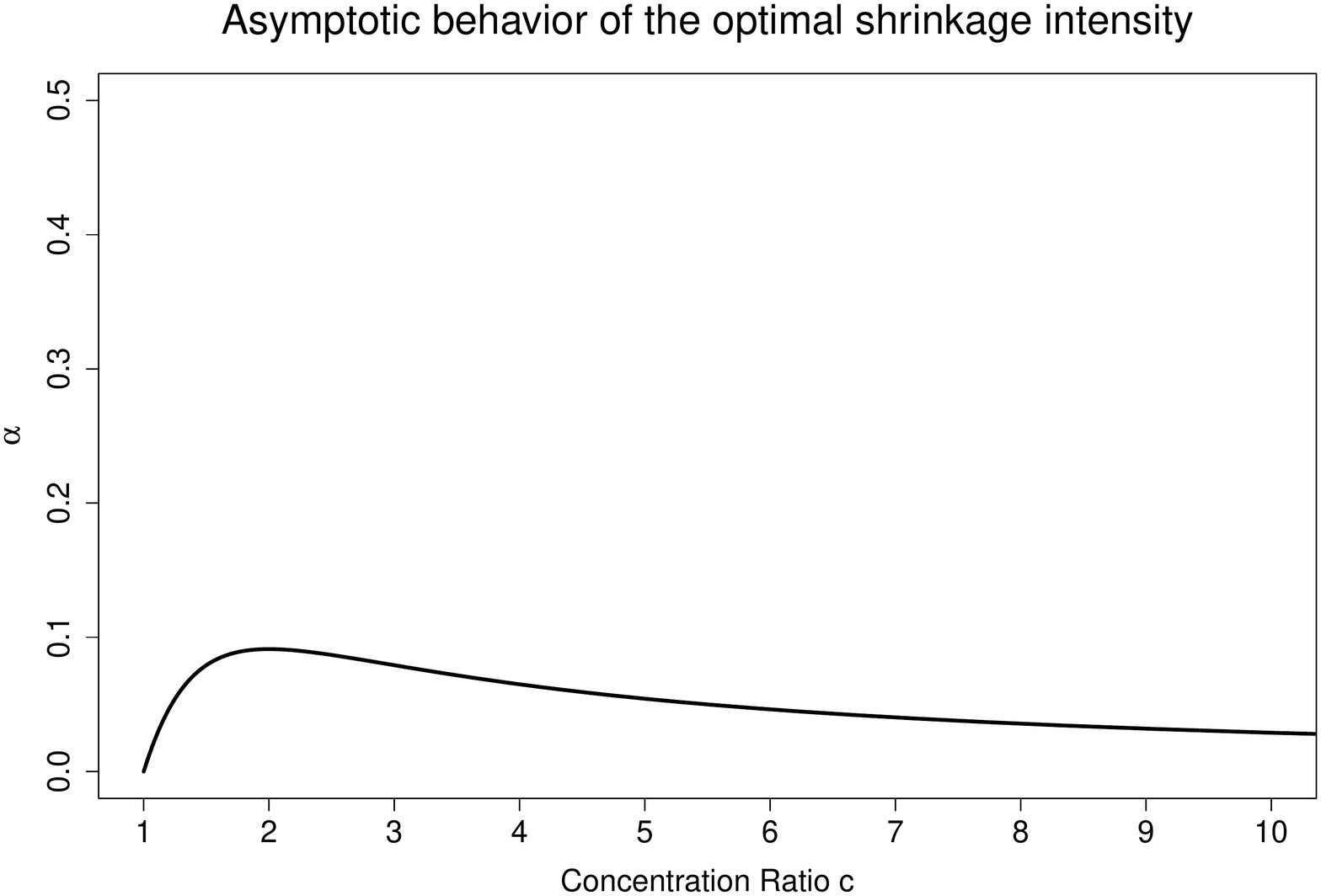}
\vspace{-5mm}
\caption{Asymptotic behavior of the optimal shrinkage intensity $\alpha^+$ as a function of the concentration ratio $p/n=c>1$.}
\label{Fig4}
\end{center}
\end{figure}

\begin{center} Figures \ref{Fig3} and \ref{Fig4} above here \end{center}

Figure \ref{Fig4} presents the asymptotic behavior of the optimal shrinkage intensity $\alpha^+$ in case $c>1$ which is no longer a monotonic function of the concentration ratio $c$ as it is observed in Figure \ref{Fig2}. The optimal shrinkage intensity attains its maximum close to $c=2$. Moreover, $\alpha^+$ remains positive even for large values of $c$, i.e. the oracle optimal shrinkage estimator converges to $\mathbf{b}_n$ for $c\to+\infty$ much slower as it was for $c \rightarrow 1_-$. On the other hand, it converges to $\mathbf{b}_n$ pretty fast for $c\to 1_+$. As a result, we don't have to expect the instability of the proposed shrinkage estimator neither in the neighborhood of $c=1$ nor for $c >> 1$.

\subsection{Estimation of unknown parameters. Bona fide estimator}\label{bonaest}
In this subsection we show how the derived oracle estimators in case $c<1$ and $c>1$, respectively, can be consistently estimated. This is achieved by estimating consistently the relative loss of the target portfolio $R_{\mathbf{b}_n}$. This result is presented in Theorem \ref{th2}.

\vspace{0.5cm}
\begin{theorem} \label{th2}
Under the assumptions (A1)-(A2) a consistent estimator of $R_{\mathbf{b}_n}$ is given by
\begin{eqnarray}\label{estR}
\text{(a)}&&\hat{R}_{\mathbf{b}_n}=(1-p/n)\mathbf{b}_n^\prime\bS_n\mathbf{b}_n\cdot\bi^\prime\bS^{-1}_n\bi-1~~\text{for} ~~ \dfrac{p}{n}\rightarrow c\in(0, 1)~~ \text{as} ~~ n\rightarrow\infty\\
\text{(b)}&&\hat{R}_{\mathbf{b}_n}^{*}=p/n(p/n-1)\mathbf{b}_n^\prime\bS_n\mathbf{b}_n\cdot\bi^\prime\bS^{*}_n\bi-1~~\text{for} ~~ \dfrac{p}{n}\rightarrow c\in(1, +\infty)~~ \text{as} ~~ n\rightarrow\infty\,.
\end{eqnarray}
\end{theorem}

\vspace{0.5cm}
The proof of Theorem \ref{th2} is given in the appendix. Applying Theorems \ref{th1} and \ref{th2}(a) allows us to determine the \textit{bona fide} estimator for the GMV portfolio weights in case $c\in(0, 1)$. It is given by
\begin{equation}\label{bonafide}
\hat{\bw}_{BFGSE}=\widehat{\alpha}^*\dfrac{\bS_n^{-1}\bi}{\bi^\prime\bS_n^{-1}\bi}+(1-\widehat{\alpha}^*)\mathbf{b}_n~~\text{with}~\widehat{\alpha}^*=
\dfrac{(1-p/n)\hat{R}_{\mathbf{b}_n}}{p/n+(1-p/n)\hat{R}_{\mathbf{b}_n}}\,,
\end{equation}
where $\hat{R}_{\mathbf{b}_n}$ is given above in (\ref{estR}). The expression (\ref{bonafide}) presents the optimal shrinkage estimator for a given target portfolio $\mathbf{b}_n$ because the shrinkage intensity $\widehat{\alpha}^*$ tends almost surely to its optimal value $\alpha^*$ for $p/n\rightarrow c\in(0, 1)$ as $n\rightarrow\infty.$

The situation is more complicated in case $c>1$. Here, the quantity $\hat{R}_{\mathbf{b}_n}$ is not a bona fide estimator of the relative loss of the target portfolio, since the matrix $\bS_n^*$ depends on unknown quantities. For that reason we propose a reasonable approximation via the the application of the Moore-Penrose inverse $\bS_n^+$. It is easy to verify that in case of $\bSigma_n=\sigma^2\bI$ equality holds. Furthermore, both the extensive simulation study of Section 3 and the empirical investigations of Section 4 document that this approximation does a very good job even for dense\footnote{opposite of sparse.} population covariance matrix $\bSigma_n$. The reason of this behavior could be the point that $\bS_n^+$ possesses a similar asymptotic behavior as $\bS_n^*$. However, it is a very challenging mathematical problem to prove this result analytically and we leave this for the future research. In Figure \ref{Fig:Moore} we provide a short simulation with the same design as presented in Figure \ref{Fig2} in order to show that $\hat{\alpha}^*(\bS_n^+)$ and $\hat{\alpha}^*(\bS_n^*)$  are close asymptotically and justify the accuracy of our approximation.

\begin{center} Figures \ref{Fig:Moore} above here \end{center}

Taking into account the above discussion and the result of Theorem \ref{th2} (b), the bona fide estimator of the quantity $R_{\mathbf{b}}$ in case $c>1$ is approximated by
\begin{equation}\label{Rb+}
\hat{R}^+_{\mathbf{b}_n} = p/n(p/n-1)\mathbf{b}_n^\prime\bS_n\mathbf{b}_n\cdot\bi^\prime\bS^{+}_n\bi-1 ~~\text{for}~~~~ c\in(1, +\infty)\,.
\end{equation}
The application of \eqref{Rb+} leads to the bona fide optimal shrinkage estimator of the GMV portfolio in case $c>1$ expressed as
\begin{equation}\label{bonafide+}
\hat{\bw}_{BFGSE}^+=\widehat{\alpha}^+\dfrac{\bS_n^{+}\bi}{\bi^\prime\bS_n^{+}\bi}+(1-\widehat{\alpha}^+)\mathbf{b}_n~~\text{with}~\widehat{\alpha}^+=
\dfrac{(p/n-1)\hat{R}^+_{\mathbf{b}_n}}{(p/n-1)^2+p/n+(p/n-1)\hat{R}^+_{\mathbf{b}_n}}\,,
\end{equation}
where $\bS_n^{+}$ is the Moore-Penrose pseudo-inverse of the sample covariance matrix $\bS_n$. 

It is noted that the estimator (\ref{bonafide}) is the optimal estimator of the GMV portfolio for $c<1$ in terms of minimizing the out-of-sample variance, while the estimator (\ref{bonafide+}) is a suboptimal one in case $c>1$. In order to summarize this section, we merge (\ref{bonafide}) and (\ref{bonafide+}) into one \textit{bona fide} optimal shrinkage estimator for the GMV portfolio weights in case $c>0$ given by\footnote{The case $c=1$ is not theoretically handled but using the Moore-Penrose inverse and setting equal to zero the smallest eigenvalue we are still able to construct a feasible estimator in this situation.}
\begin{equation}\label{Bonafide}
\hat{\bw}_{BFGSE}=\widehat{\alpha}^*\dfrac{\bS_n^{+}\bi}{\bi^\prime\bS_n^{+}\bi}+(1-\widehat{\alpha}^*)\mathbf{b}_n~~\text{with}\,
\end{equation}
\begin{equation}\label{conestal}
\widehat{\alpha}^*=\left\{
\begin{array}{lr}
\dfrac{(1-p/n)\hat{R}_{\mathbf{b}_n}}{p/n+(1-p/n)\hat{R}_{\mathbf{b}_n}}&~~\text{for}~c<1,\\
\dfrac{(p/n-1)\hat{R}_{\mathbf{b}_n}}{(p/n-1)^2+p/n+(p/n-1)\hat{R}_{\mathbf{b}_n}}&~~\text{for}~c\geq1\,,
\end{array}
\right.
\end{equation}
and
\begin{equation}\label{rbona}
\hat{R}_{\mathbf{b}_n}=\left\{
\begin{array}{lr}
(1-p/n)\mathbf{b}_n^\prime\bS_n\mathbf{b}_n\cdot\bi^\prime\bS^{-1}_n\bi-1&~~\text{for}~c<1,\\
p/n(p/n-1)\mathbf{b}_n^\prime\bS_n\mathbf{b}_n\cdot\bi^\prime\bS^{+}_n\bi-1&~~\text{for}~c\geq1\,.
\end{array}
\right. \,,
\end{equation}
where we use that $\bS_n^+=\bS_n^{-1}$ if $\bS_n$ is nonsingular.

\begin{figure}[h!tb]
\begin{center}
\includegraphics[scale=0.4]{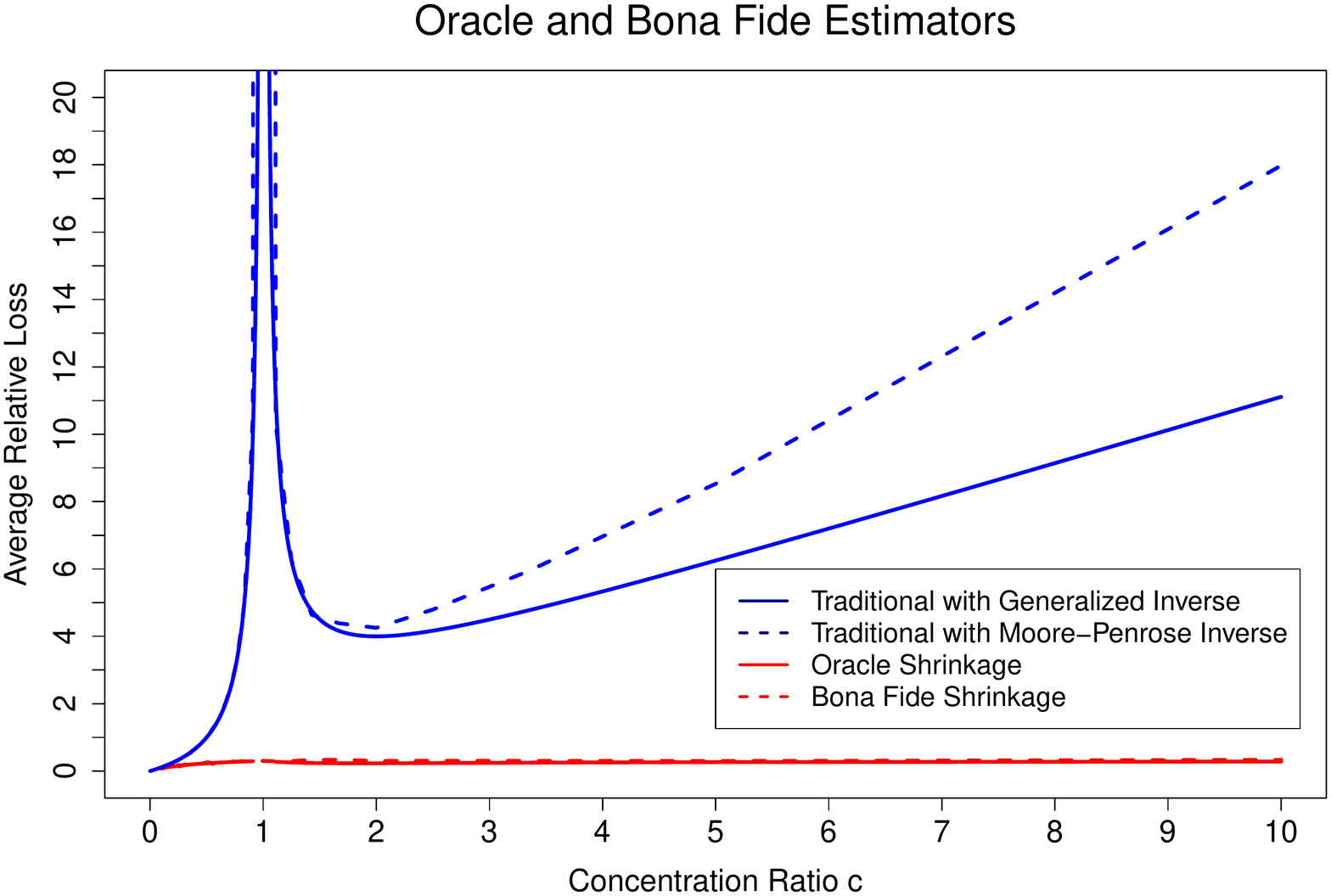}
\caption{Oracle and bona fide traditional and optimal shrinkage estimators of the GMV portfolio for different values of $p/n=c>0$.}
\label{Fig5}
\end{center}
\end{figure}

\begin{landscape}
\begin{figure}[h!tb]
\center\scalebox{0.8}{
\includegraphics[]{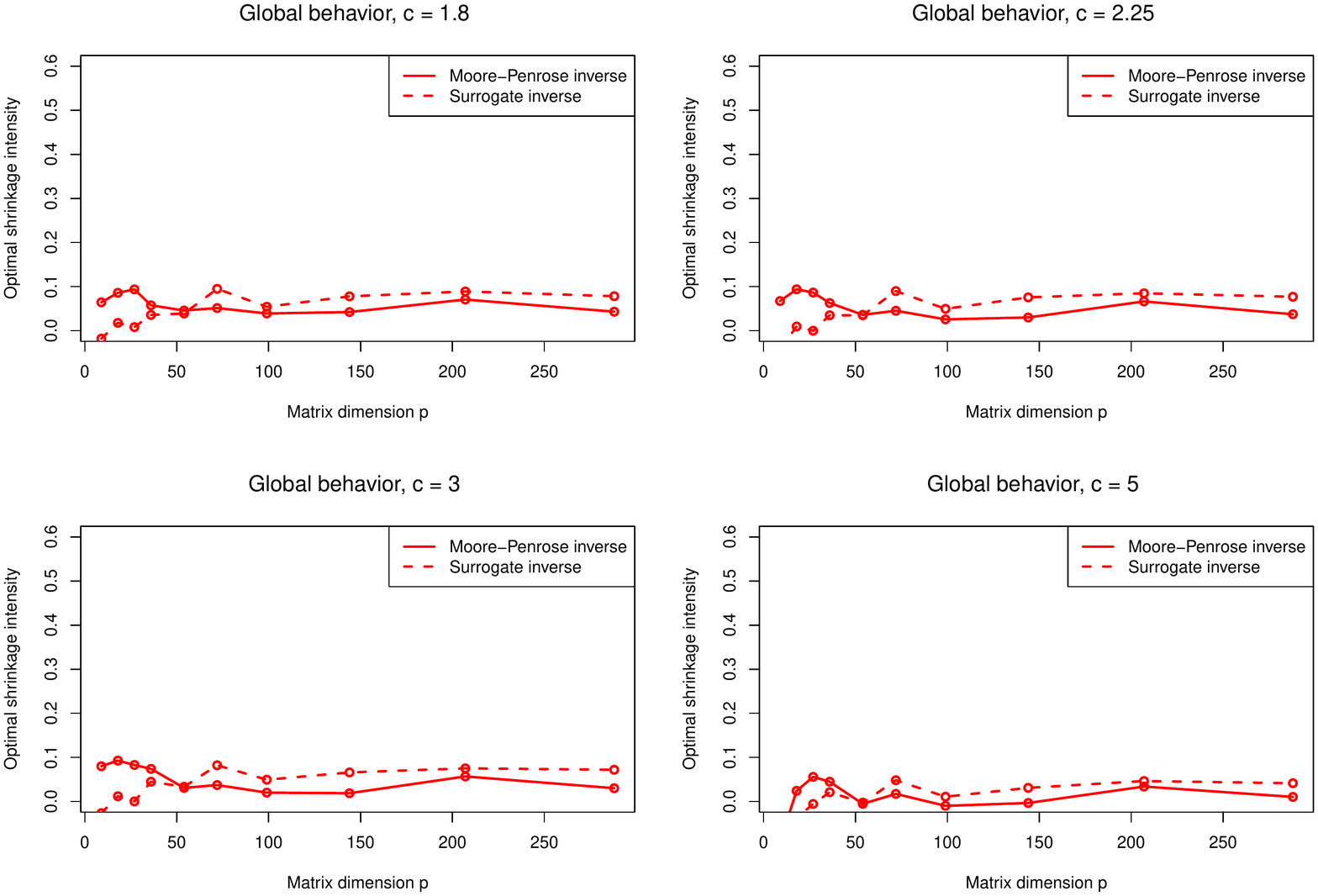}}
\vspace{-15mm}
\caption{Simulation results on the accuracy of approximation for normally distributed data in case of the bounded spectrum ($c=\{1.8, 2.55, 3, 5\}$, 1000 repetitions).}
\label{Fig:Moore}
\end{figure}
\end{landscape}

In Figure \ref{Fig5} we investigate the difference between the oracle and the bona fide optimal shrinkage estimators for the GMV portfolio weights as well as between the oracle and the bona fide traditional estimators. The population covariance matrix is taken as a dense $207\times207$-dimensional covariance matrix $\bSigma_n$ with $1/9$ of eigenvalues equal to $2$, $4/9$ to $5$, and last $4/9$ to $10$. The eigenvectors are chosen in the same way as in the section about oracle estimator. The target portfolio is still the naive one, i.e., $\mathbf{b}_n=1/p\bi$. The observation matrix is generated from the normal distribution.

A perfect fit of the bona fide optimal shrinkage estimator (dotted red line) to its oracle (solid red line) is observed for all of the considered values $c>0$. The blue lines corresponds to the oracle traditional estimator (solid blue line) and the bona fide traditional estimator (dash blue line). In contrast to the optimal shrinkage estimator, a difference between the bona fide traditional estimator and its oracle is present for $c>1$ which increases as $c$ becomes larger. For $c<1$ both the estimators coincide since in this case both the generalized inverse \eqref{geninverse} and the Moore-Penrose inverse are equal to the inverse of the sample covariance matrix. It is remarkable that the proposed bona fide optimal shrinkage estimator works well also at point $c=1$ although the corresponding oracle estimator is even not defined there. The reason is that we just set equal to zero the smallest eigenvalue of $\bS_n$ and use the Moore-Penrose inverse technique. The results of Figure \ref{Fig5} motivate the application of the Moore-Penrose inverse instead of the generalized inverse given at the beginning of Section 2.2 in practice, whereas the traditional estimator should be used with care. We provide a further investigation of this point in the simulation study of Section 3.

\begin{center} Figure \ref{Fig5} above here \end{center}

The last point, which has to be noted, is that the bona fide estimator (\ref{Bonafide}) is easy to use in practice since it can be fast computed.

\subsection{Choice of the target portfolio}

The target portfolio $\mathbf{b}_n$ plays a crucial role in the determination of the optimal shrinkage estimator. The most obvious choice of $\mathbf{b}_n$ would be the naive portfolio $\frac{1}{p}\bi$ or a sparse portfolio. In the multi-period setting the weights of the previous period can be chosen as a target portfolio. Theoretically, we can even take a random target portfolio but it should be independent of the actual observations. In particular, it can be a uniformly distributed random vector on the unit sphere (suitably normalized) or a uniformly distributed random vector on the simplex. Choosing the optimal portfolio weights of the previous periods leads to more interesting example for a target portfolio which allows us to construct some sort of Bayesian updating principle in the dynamic setting.

In general, the answer on this question depends on the underlying data because the choice of the target weights is equivalent to the choice of the hyperparameter for the prior distribution of $\frac{\bSigma^{-1}_n\bi}{\bi^\prime\bSigma_n^{-1}\bi}$. This problem is well-known in Bayesian statistics. The application of different priors leads to different results. So it is very important to choose the one which works well in most cases. The naive one is the equally weighted portfolio $1/p\bi$. Obviously, the oracle shrinkage estimator with the prior weights as the true global minimum variance portfolio is a consistent estimator as shown in Proposition \ref{prior_prop}. Moreover, including some new information about the true GMV portfolio into the prior can lead to a significant increase of performance (see, Bodnar et al. (2014)). For simplicity we take the naive portfolio in our simulation study in Section 3 as well as in the empirical investigation of Section 4.

Consider the shinkage estimator as a vector function $\hat{\bw}_{BFGSE}(\mathbf{b}_n): V_p \rightarrow \tilde{V}_p$, where $V_p$ and $\tilde{V}_p$ are the $p$-dimensional vector spaces. In the following proposition we present some properties of the shrinkage estimator as a function of the target weights $\mathbf{b}_n$.
\begin{proposition}\label{prior_prop}
For the proposed shrinkage estimator $\hat{\bw}_{BFGSE}(\mathbf{b}_n)$ it holds that
\begin{enumerate}
\item $\hat{\bw}_{BFGSE}(1/p\bi)$ is a consistent estimator for the GMV portfolio if the population covariance matrix  $\bSigma_n=\sigma\bI$ for arbitrary $\sigma>0$ and $c\in(0, +\infty)$.
\vspace{3mm}
\item $\hat{\bw}_{BFGSE}(\bw_{GMV})$ is a consistent estimator for the GMV portfolio $\frac{\bSigma_n^{-1}\bi}{\bi^\prime\bSigma_n^{-1}\bi}$ for all $c\in(0, +\infty)$.
\end{enumerate}
\end{proposition}
The proof is a straightforward application of Theorem \ref{th1}, Theorem \ref{th12} and Theorem \ref{th2}.

\section{Simulation study}

In this section we demonstrate how the obtained results can be applied in practice. The first part of our simulations is dedicated to normally distributed data, while in the second part the asset returns are generated from the $t$-distribution with $5$ degrees of freedom. The target portfolio $\mathbf{b}_n$ is taken as the naive portfolio $\dfrac{\bi}{p}$. The results are presented in both cases $c<1$ and $c>1$ as well as for the covariance matrix with bounded (Section 3.1) and unbounded (Section 3.2) spectrum.

The benchmark estimator is the dominating estimator of the GMV portfolio suggested by Frahm and Memmel (2010). It is given by
\begin{equation}\label{FM}
\hat{\bw}_{FM}=(1-k)\dfrac{\bS_n^{-1}\bi^\prime}{\bi^\prime\bS_n^{-1}\bi}+k\dfrac{\bi}{p}\,~\text{with}~~k=\dfrac{p-3}{n-p+2}\dfrac{1}{\hat{R}_{\bi/p}}\,,
\end{equation}
where $\hat{R}_{\bi/p}=\dfrac{1/p^2\bi^\prime\bS_n\bi-\sigma^2_{\bS_n}}{\sigma^2_{\bS_n}}$ is the estimated relative loss of the naive portfolio. The dominating estimator (\ref{FM}) is derived under the assumption that the asset returns are normally distributed and it dominates over the traditional estimator in terms of the out-of-sample variance (cf. Frahm and Memmel (2010)). Nevertheless, it is not clear how far it is away from the optimal one for different values of the concentration ratio $c>0$. Its behavior for non-normally distributed data has not been studied yet as well.

Next, we compare the performance of the dominating estimator (\ref{FM}) with the bona fide optimal shrinkage estimator (\ref{Bonafide}). In order to find out the rates of convergence established in Theorem \ref{th1} and \ref{th2}, we also consider the oracle optimal shrinkage estimator which can be easily constructed for $c<1$ and $c>1$ with the optimal shrinkage intensities given by (\ref{alfa}) and (\ref{alfa+}), respectively. As a performance measure we take the relative loss from Section 2. For an arbitrary estimator $\hat{\bw}$ of the GMV portfolio it is defined by
\begin{equation}\label{perf}
R_{\hat{\bw}}=\dfrac{\sigma^2_{\hat{\bw}}-\sigma^2_{GMV}}{\sigma^2_{GMV}}\,
\end{equation}
where $\sigma^2_{\hat{\bw}}=\hat{\bw}^\prime\bSigma_n\hat{\bw}$ and $\sigma^2_{GMV}=\dfrac{1}{\bi^\prime\bSigma^{-1}_n\bi}$.

In our simulation study we take $p$ as a function of $n$. In particular, when $n=18\cdot2^j$ and $p=9\cdot2^j$ for $j\in[0, 5]$ the concentration ration $c$ is always equal to $0.5$ and $p$ increases together with $n$ exponentially. That is why the small dimensions are presented with more points and the large ones with less. Similar choices of $p$ and $n$ are also performed for other values of $c \in \{0.1,0.9,1.8\}$. Finally, it is noted that the simulation results show a good convergent rate in terms of the relative loss for the bona fide optimal shrinkage estimator to its oracle one already for $p\le100$.

\subsection{Population covariance matrix with bounded spectrum}
In this subsection, we assume that the covariance matrix possesses a bounded spectrum, i.e. with bounded maximum eigenvalue. Here, we use the structure of the covariance matrix as in Figure \ref{Fig5}, i.e., we take $1/9$ of its eigenvalues equal to $2$, $4/9$ equal to $5$ and $4/9$ equal to $10$. The high-dimensional covariance matrices constructed in this way possess uniformly the same spectral norm and their eigenvalues are not very dispersed. 
Additionally, this choice of the covariance matrices ensures that when the dimension $p$ increases then the spectrum of the covariance matrices does not change its behavior.

\begin{landscape}
\begin{figure}[h!tb]
\center\scalebox{0.8}{
\includegraphics[]{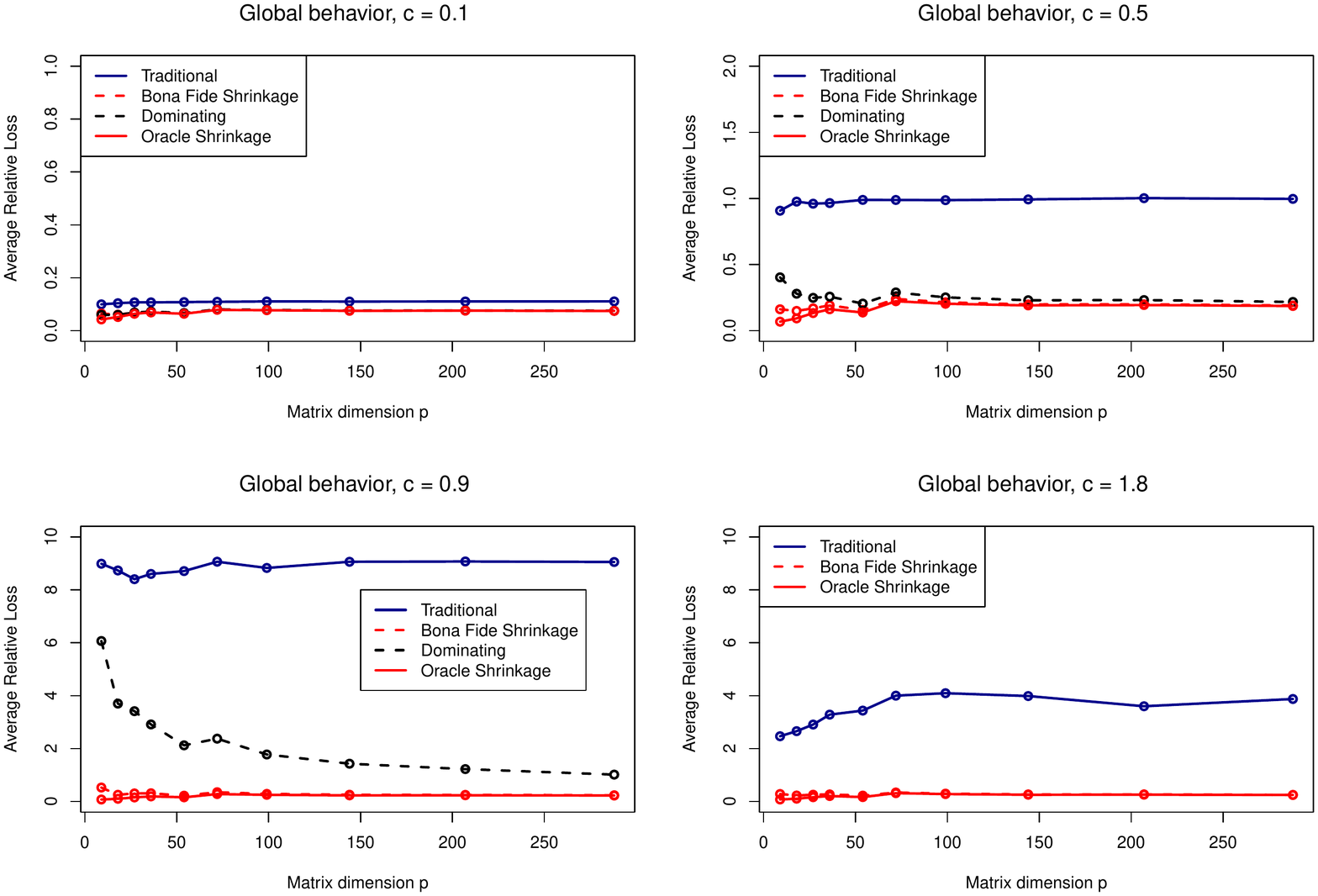}}
\vspace{-15mm}
\caption{Simulation results for normally distributed data in case of the bounded spectrum ($c=\{0.1, 0.5, 0.9, 1.8\}$, 1000 repetitions).}
\label{Fig:6}
\end{figure}
\end{landscape}

\begin{landscape}
\begin{figure}[h!tb]
\center\scalebox{0.8}{
\includegraphics[]{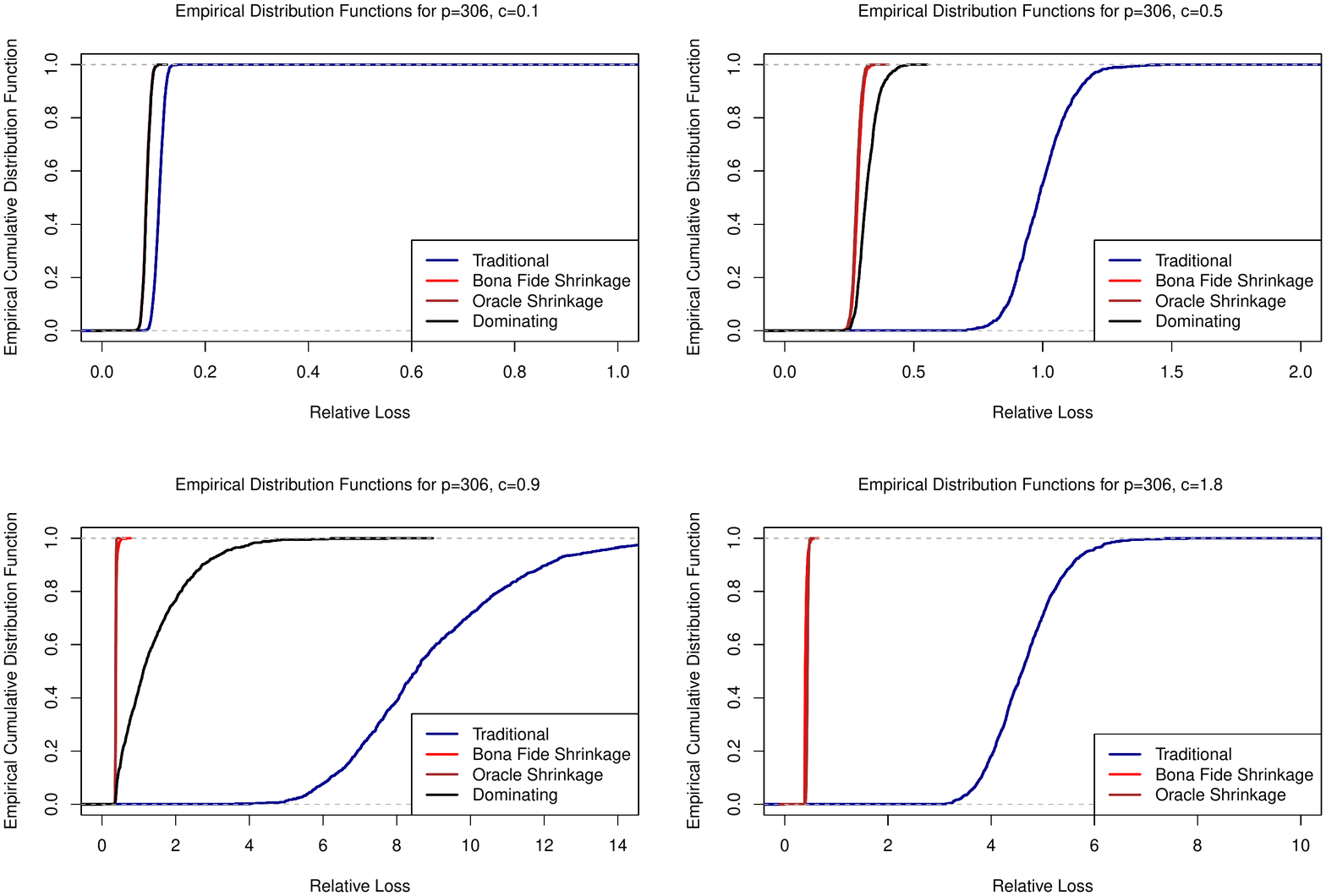}}
\vspace{-15mm}
\caption{Simulation results for normally distributed data in case of the bounded spectrum ($c=\{0.1, 0.5, 0.9, 1.8\}$, 1000 repetitions).}
\label{Fig:7}
\end{figure}
\end{landscape}

\begin{center} Figures \ref{Fig:6} and \ref{Fig:7} above here \end{center}

In Figures \ref{Fig:6} and \ref{Fig:7} we present the simulation results for normally distributed data and different values of the concentration ratio $c\in\{0.1,0.5, 0.9, 1.8\}$. Figure \ref{Fig:6} presents the \textit{global} behavior of the considered estimators for different dimensions $p$, while Figure \ref{Fig:7} shows the \textit{local} distributional properties for a value of fixed $p=306$. More precisely, under the \textit{global} behavior we understand the evolution of the average relative loss with respect to the dimension $p$ and the \textit{local} behavior presents the empirical cumulative distribution functions (e.c.d.f.) of the relative loss for one fixed value of $p$, namely $p=306$. The comparison in case of \textit{global} setting is clear: the smaller the average loss the better is the estimator. The \textit{local} study provides a more precise comparison in terms of the empirical distributions. In this case, the criterion of the best estimator is based on the observation that the e.c.d.f. with stochastically smaller values is dominating. This means that for two e.c.d. functions, the dominating one is placed on the left side from the other. This criterion is consistent with the stochastic dominance of order one. The only difference with respect to the stochastic dominance of order one is that the comparison is based on the empirical distribution functions instead of the population one.


On global analysis we see that the bona fide optimal shrinkage estimator converges to the corresponding oracle one already for small values of $p$ in all of the considered cases $c \in \{0.1,0.5, 0.9, 1.8\}$. 
On the third place, the dominating estimator of Frahm and Memmel (2010) is ranked. It is always better than the traditional estimator which is the worst one, but it is always worse than the other two competitors. In terms of the values of the average relative loss, we observe that the difference between the estimators become more significant if $c$ increases and lies below $1.0$. For instance, in case of $c=0.1$, the average relative loss of the traditional estimator tends to $1/9$, whereas it tends to $1$ for $c=0.5$. These two results are in line with Corollary \ref{cor1}, where it is proved that the average loss of the traditional estimator tends to $c/(1-c)$ under high-dimensional asymptotics. In case of $c=0.9$, the difference between the average relative loss of the optimal shrinkage estimator and the dominating (traditional) estimators becomes very large. Indeed, in this case the traditional estimator has an average relative loss which is asymptotically equal to $9$. This means that the out-of-sample risk of the traditional estimator is $10$ times larger than the real risk. The dominating estimator clearly overperforms the traditional one but the relative loss is close to $4$ for small dimensions ($p\leq50$) which means that its out-of-sample risk is $5$ times as large as the real risk. This is not acceptable anymore. In contrast, the \textit{bona fide} optimal shrinkage estimator converges fast to its \textit{oracle} one. The relative loss of the optimal shrinkage estimator is smaller than $0.3$.

Figure \ref{Fig:7} shows the same dominance in terms of the empirical distribution functions for a local analysis in the case $p=306$. The best approaches are the \textit{oracle} and the \textit{bona fide} optimal shrinkage estimators. Next, the dominating estimator is ranked followed by the traditional one. The plots also illustrate the fast convergence of the \textit{bona fide} optimal shrinkage estimator to its oracle. The local analysis for $p=306$ confirms the almost sure convergence (consistence) of the \textit{bona fide} optimal shrinkage estimator which is proved in Theorem \ref{th1}. In Figure \ref{Fig:7} the relative risk of both the bona fide and the oracle optimal shrinkage estimators possesses a very small variance which vanishes when the dimension $p$ increases. At the same time, the dominating estimator possesses a significantly larger variance and it is unstable when $c$ is close to one. The traditional estimator shows a very crucial behavior and it is the worst one among the considered estimators.

The most interesting situation is observed for $c=1.8$ in Figures \ref{Fig:6} and \ref{Fig:7} which corresponds to the singular sample covariance matrix $\bS_n$. Here, we apply the results from Section \ref{secc1} and \ref{bonaest} and take the Moore-Penrose inverse $\bS_n^+$ instead of $\bS_n^{-1}$. Note that we cannot use the dominating estimator because it is not applicable for $c>1$. The results are still impressing for both the global and the local regimes. Again, the proposed bona fide optimal shrinkage estimator converges to its oracle. As a traditional estimator, we take the GMV portfolio constructed by using the Moore-Penrose inverse $\bS_n^+$. The traditional estimator possesses a rapidly increasing average loss and the largest variance. It is not an acceptable estimator also for $c>1$. In contrast, the bona fide optimal shrinkage estimator has a small variance and obeys a stable behavior even if $c>1$.

Further we analyze the behavior of the considered estimators when the asset returns are no longer normally distributed. In particular it is interesting to study how strong is the impact of heavy tails on the estimators derived in the paper. For this reason, the $t$-distribution with $5$ degrees of freedom is used next in our simulation study. Recently, authors have mentioned that $5$ degrees of freedom seems to be a suitable choice in practice (see, Venables and Ripley (2002)).

\begin{landscape}
\begin{figure}[h!tb]
\center\scalebox{0.8}{
\includegraphics[]{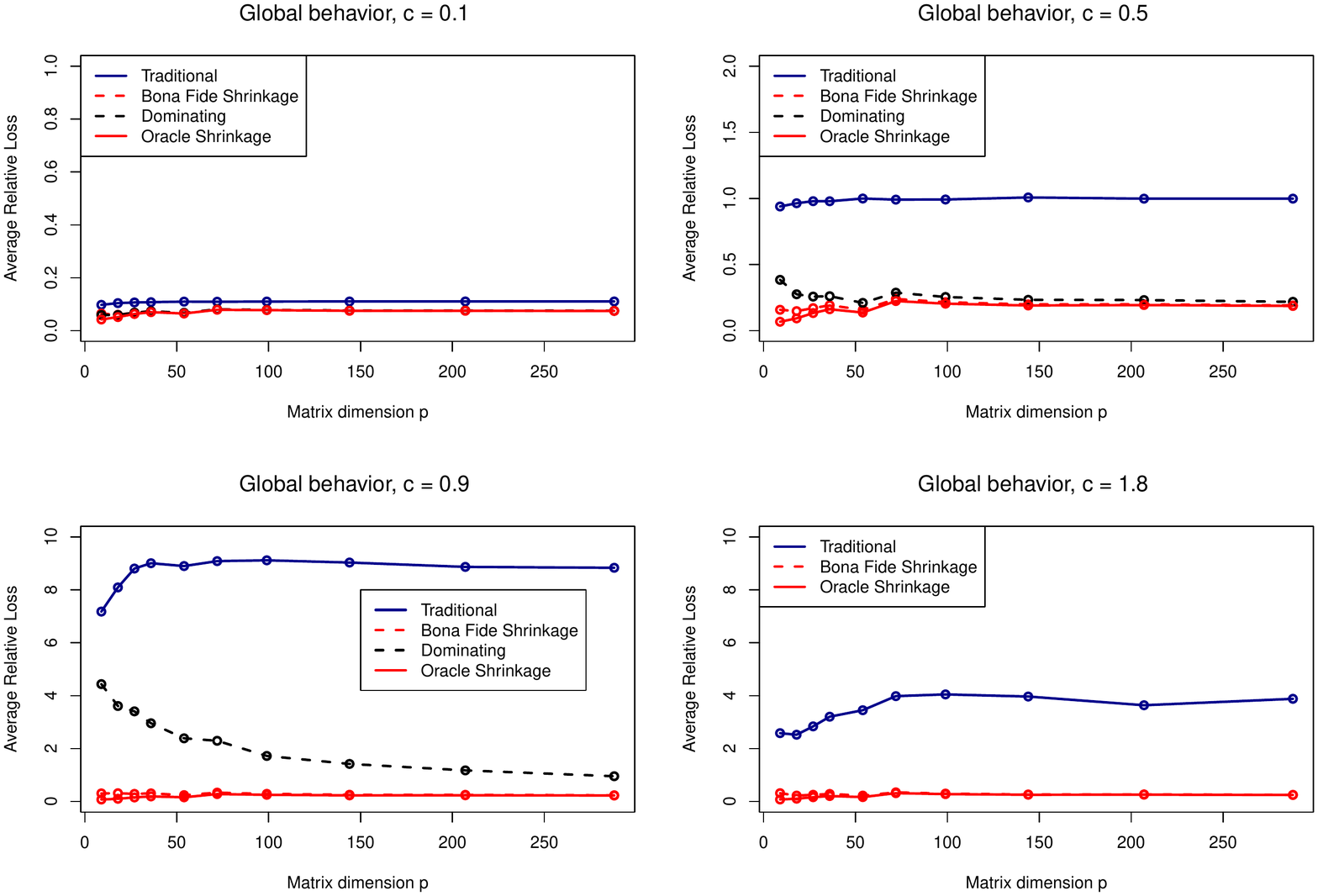}}
\vspace{-15mm}
\caption{Simulation results for $t$-distributed data with $5$ degrees of freedom in case of the bounded spectrum ($c=\{0.1, 0.5, 0.9, 1.8\}$, 1000 repetitions).}
\label{Fig:8}
\end{figure}
\end{landscape}

\begin{landscape}
\begin{figure}[h!tb]
\center\scalebox{0.8}{
\includegraphics[]{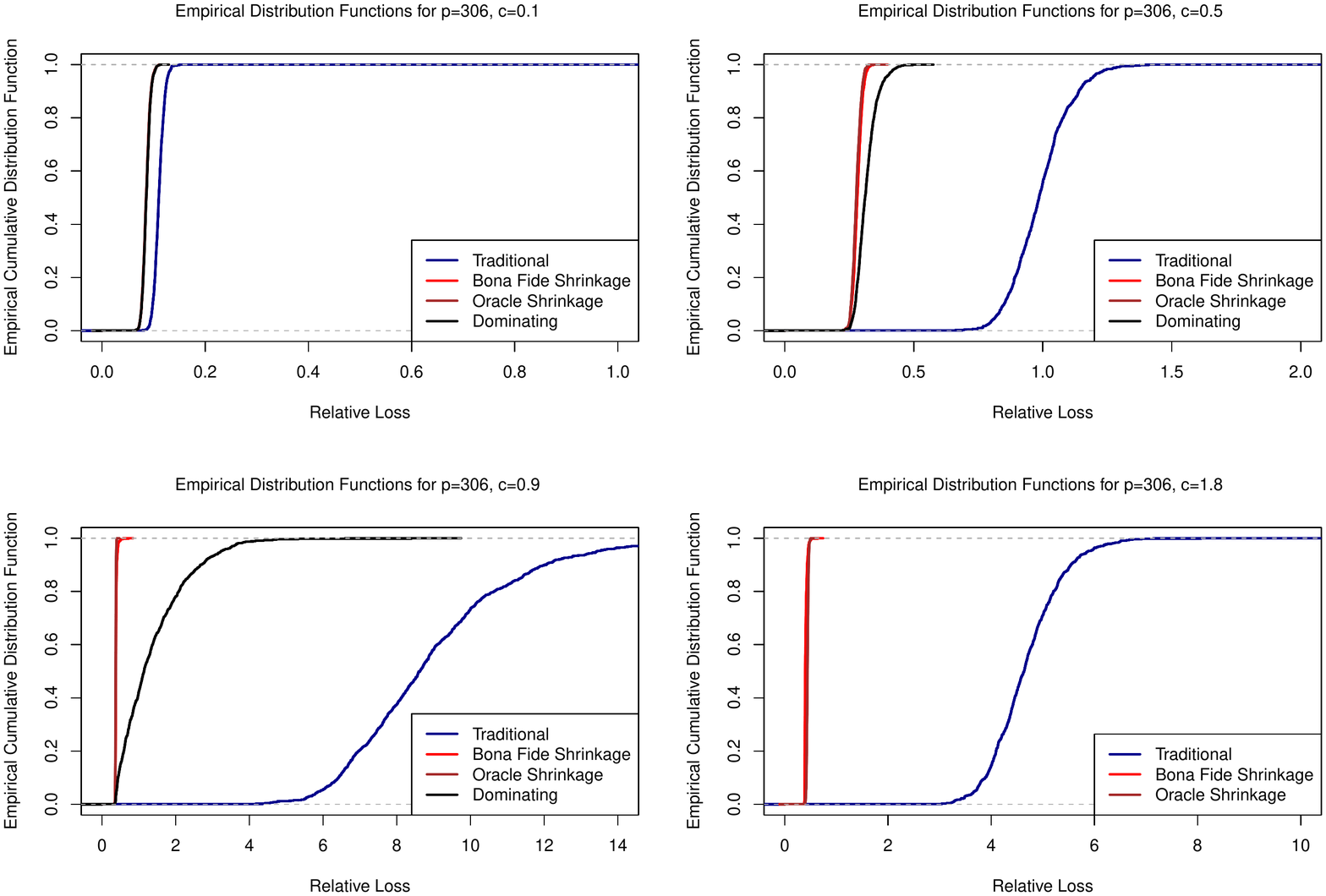}}
\vspace{-15mm}
\caption{Simulation results for $t$-distributed data with 5 degrees of freedom in case of the bounded spectrum ($c=\{0.1, 0.5, 0.9, 1.8\}$, 1000 repetitions).}
\label{Fig:9}
\end{figure}
\end{landscape}

\begin{center} Figures \ref{Fig:8} and \ref{Fig:9} above here \end{center}

In Figures \ref{Fig:8} and \ref{Fig:9} we present the results for the $t$-distributed asset returns with $5$ degrees of freedom. The structure of the comparison study is the same as in case of normally distributed data. In general, the behavior observed in Figures \ref{Fig:8} and \ref{Fig:9} does not differ significantly from those obtained for the normal distribution. The best estimator is, as usual, the proposed shrinkage estimator. The optimal shrinkage estimator dominates clearly other competitors over all $c\in\{0.1,0.5, 0.9, 1.8\}$. It is noted that the convergence rate of the \textit{bona fide} optimal shrinkage estimator to its \textit{oracle} is not effected by the presence of heavy tails. A similar asymptotic relative loss behavior for the optimal shrinkage estimator is established, i.e., the average relative loss is asymptotically constant and it is smaller than $0.5$. The traditional estimator possesses the worst behavior over all $c$ and $p$.

\subsection{Population covariance matrix with unbounded spectrum}
In this subsection we assume that the largest eigenvalue of the population covariance matrix $\bSigma_n$ increases as $O(p)$ when $p\rightarrow\infty$. Thus, the following structure of $\bSigma_n$ is considered here, namely $1/9$ of eigenvalues equal to $2$, $4/9$ equal to $5$, $(4/9p-1)$ equal to $10$ and the last eigenvalue is equal to $p$.\\ Note that this structure corresponds to the case when a factor structure on the asset returns is imposed. The factor model can reduce significantly the number of dimensions so that the estimators do not suffer from the "curse of dimensionality" anymore (see, e.g., Fan et al. (2013)).

\begin{landscape}
\begin{figure}[h!tb]
\center\scalebox{0.8}{
\includegraphics[]{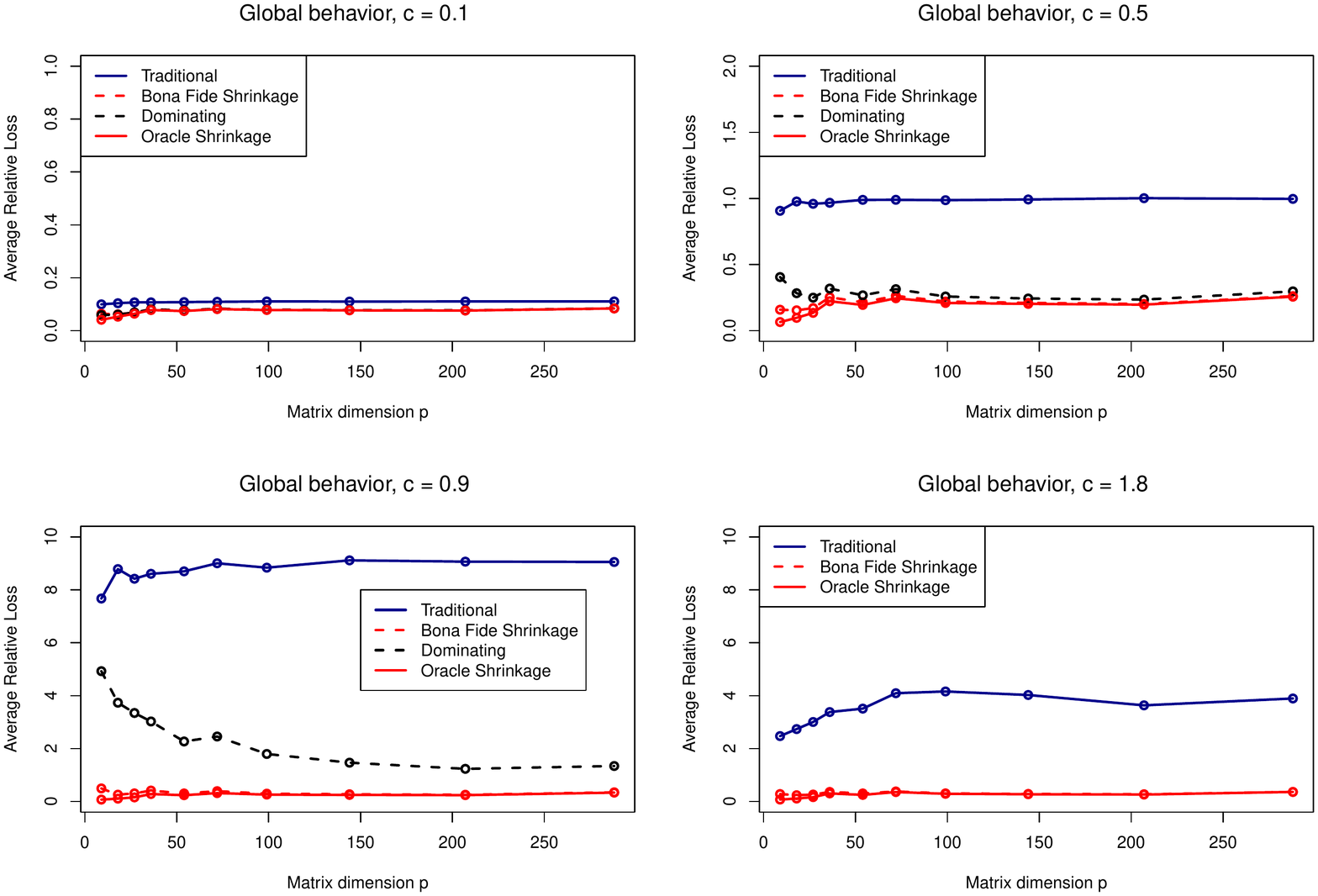}}
\vspace{-15mm}
\caption{Simulation results for normally distributed data in case of the unbounded spectrum ($c=\{0.1, 0.5, 0.9, 1.8\}$, 1000 repetitions).}
\label{Fig:10}
\end{figure}
\end{landscape}

\begin{landscape}
\begin{figure}[h!tb]
\center\scalebox{0.8}{
\includegraphics[]{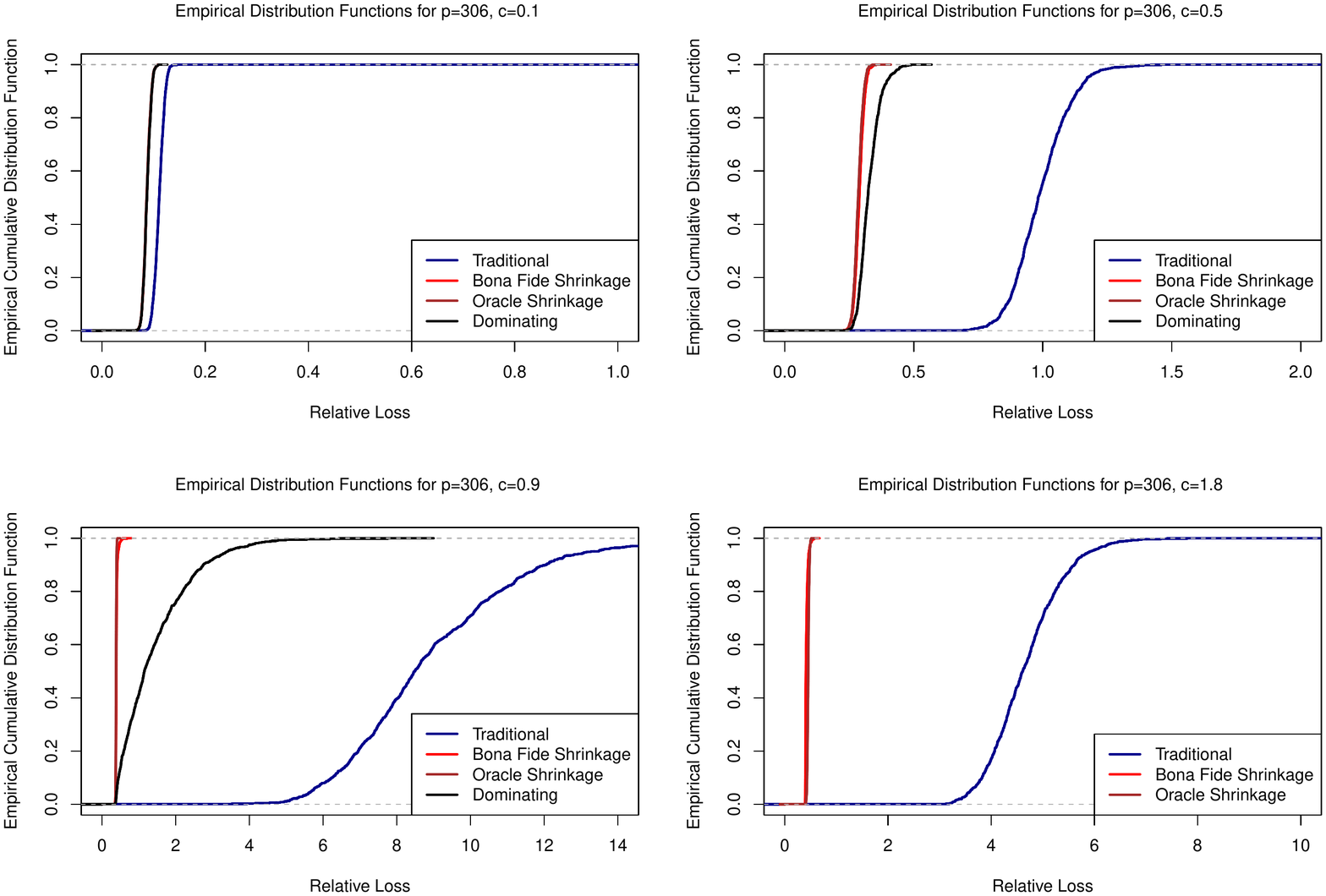}}
\vspace{-15mm}
\caption{Simulation results for normally distributed data in case of the unbounded spectrum ($c=\{0.1, 0.5, 0.9, 1.8\}$, 1000 repetitions).}
\label{Fig:11}
\end{figure}
\end{landscape}

\begin{landscape}
\begin{figure}[h!tb]
\center\scalebox{0.8}{
\includegraphics[]{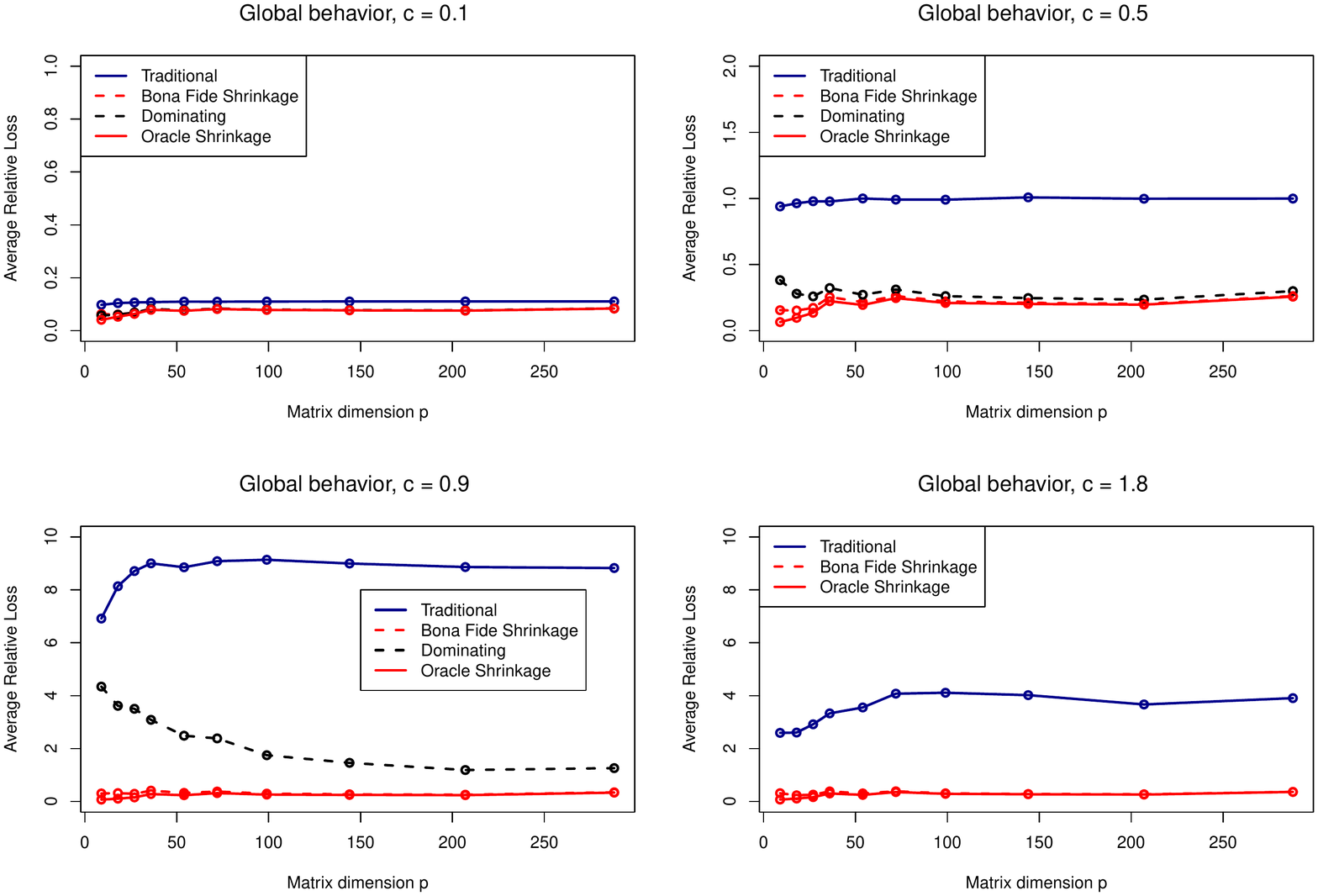}}
\vspace{-15mm}
\caption{Simulation results for $t$-distributed data with $5$ degrees of freedom in case of the unbounded spectrum ($c=\{0.1, 0.5, 0.9, 1.8\}$, 1000 repetitions).}
\label{Fig:12}
\end{figure}
\end{landscape}

\begin{landscape}
\begin{figure}[h!tb]
\center\scalebox{0.8}{
\includegraphics[]{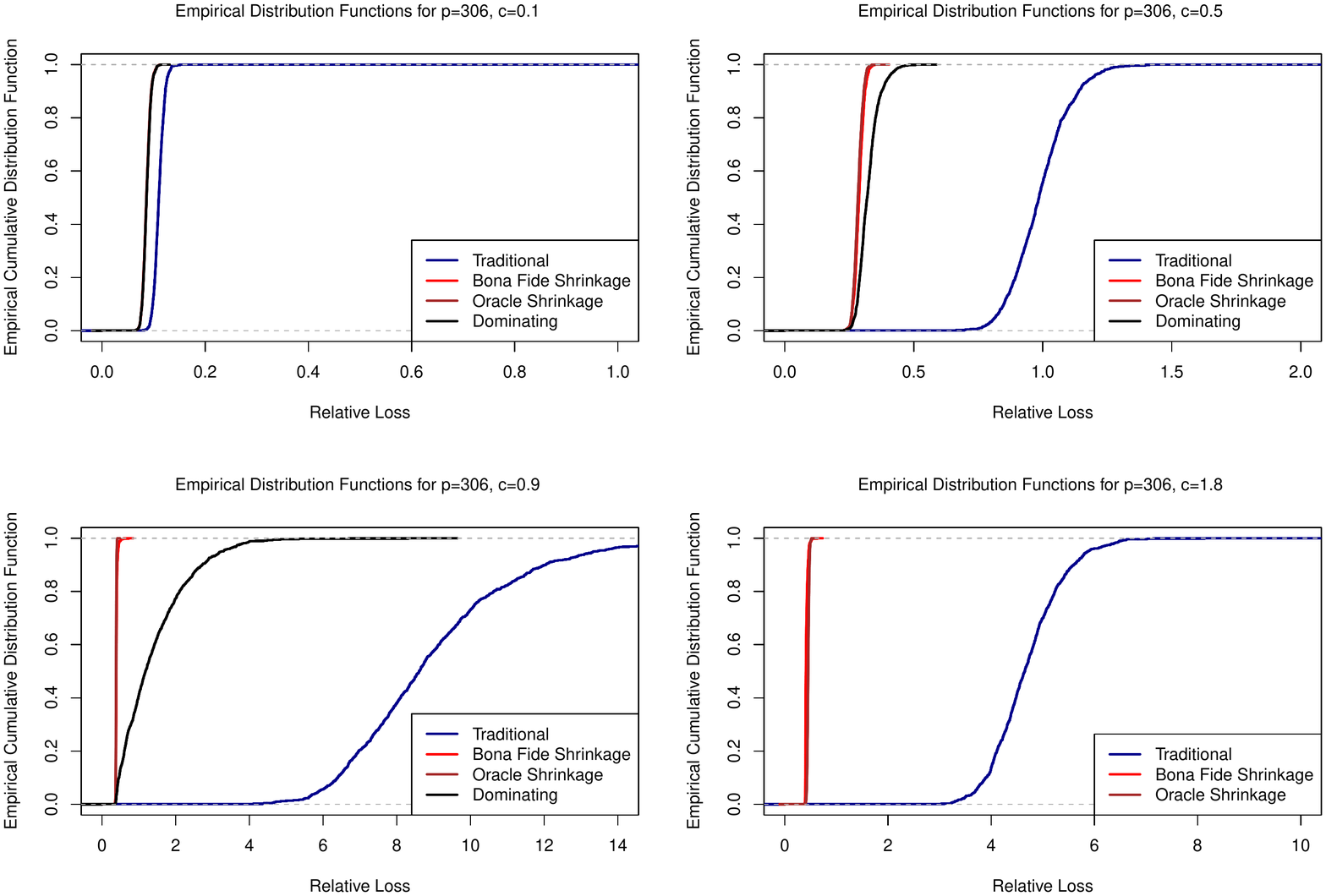}}
\vspace{-15mm}
\caption{Simulation results for $t$-distributed data with 5 degrees of freedom in case of the unbounded spectrum ($c=\{0.1, 0.5, 0.9, 1.8\}$, 1000 repetitions).}
\label{Fig:13}
\end{figure}
\end{landscape}

\begin{center} Figures \ref{Fig:10} to \ref{Fig:13} above here \end{center}

In Figures \ref{Fig:10} to \ref{Fig:13} we present the behavior of the  estimators considered in the paper in case of a covariance matrix with unbounded spectrum. It is remarkable to note that the results are not very different from those obtained in case of a covariance matrix $\bSigma_n$ with bounded spectrum. The only difference is a somewhat greater variance of the estimators. On the other hand, the dominance behavior as well as the convergence rate of the bona fide optimal shrinkage estimator to its oracle is not effected by the largest eigenvalue of the population covariance matrix. This means that the proposed estimator is still applicable if the asset returns follow a factor model. Even more, it does not lose its efficiency also in case $c>1$.

At the end, we note that, for the sake of interest, we have also simulated the $t$-distribution with $3$ degrees of freedom for both the bounded and the unbounded spectra. This change effects only the convergence rate but not the dominance behavior. In our theoretical framework we require the existence of the $4$th moment but the simulation study shows that this assumption can be relaxed or conjectured to be relaxed. As a result, the proposed optimal shrinkage procedure assures the efficiency in many important practical cases and, thus, can be applied in many real life situations. Nevertheless, the empirical back-testing is still needed in order to check the behavior of the derived estimator for the GMV portfolio weights on a real data set. This is done is the next section.

\section{Empirical Study}
In this section, we apply the proposed optimal shrinkage estimator for the GMV portfolio (\ref{Bonafide}) to real data which consist of daily returns on the 417 assets listed in the S$\&$P $500$ (Standard $\&$ Poor's $500$) index and traded during the period from 22.04.2013 to 19.03.2014. It corresponds to the horizon of $T=230$ trading days. The S\&P 500 index is based on the market capitalizations of 500 large companies having common stock listed on the NASDAQ.

In this empirical study we compare the performance of the derived optimal shrinkage estimator for the GMV portfolio weights given by (\ref{Bonafide}) with the traditional estimator and the dominating estimator suggested by Frahm and Memmel (2010). The comparison is based a procedure which is similar to the rolling-window approach proposed by DeMiguel et al. (2009). In particular, we randomly pick up a portfolio of dimension $p=54$ from all $417$ portfolios and estimate the portfolio weights for the given estimation window of the length $n<T$. We repeat this rolling-window procedure for the next step by including data of the next day and dropping out the data of the last day until the end of the data set is reached.
The estimation window $n$ is chosen such that the concentration ratio $c=p/n$ lies in the set $\{0.5, 0.9, 1.5, 2\}$.

In order to compare the performance of the estimators we consider the out-of-sample variance and the out-of-sample Sharpe ratio. Let $\hat{\bw}_t$ be an estimator for the GMV portfolio which is based on the window with last observation at time $t$ and let $\mathbf{r}_{t+1}$ be the vector of the asset returns for the next period $t+1$. Then the out-of-sample variance and the out-of-sample Sharpe ratio are calculated by
\begin{align}
&\hat{\sigma}_{out}^2=\dfrac{1}{T-n-1}\sum\limits_{t=n}^{T-1}(\hat{\bw}_t^\prime \mathbf{r}_{t+1}-\hat{\mu}_t)^2~~\text{and}~~\widehat{SR}=\dfrac{\hat{\mu}_t}{\hat{\sigma}_{out}} ~~ \text{with} ~~ \hat{\mu}_t=\dfrac{1}{T-n}\sum\limits_{t=n}^{T-1}\hat{\bw}_t^\prime \mathbf{r}_{t+1}\,.
\end{align}

In order to measure the statistical significance we sample randomly $1000$ different portfolios and calculate the e.c.d. functions of their out-of-sample variances and the corresponding out-of-sample Sharpe ratios. The best strategy is chosen similarly to the stochastic dominance principle, i.e. we look for the strategy whose e.c.d.f. stochastically dominates the other ones. However, the dominance is defined differently in case of the out-of-sample variance and the out-of-sample Sharpe ratio. For the out-of-sample variance, the e.c.d.f. of the best strategy should lie above the e.c.d. functions of the other competitors, i.e. larger values of the out-of-sample variance can take place with smaller probability. In contrast, criteria based on the out-of-sample Sharpe ratio  prefers the strategy whose e.c.d.f. lies below the other e.c.d. functions. In this case, the GMV portfolio constructed using the corresponding estimator would possess the highest out-of-sample Sharpe ratio.

\begin{landscape}
\begin{figure}[h!tb]
\center\scalebox{0.8}{
\includegraphics[]{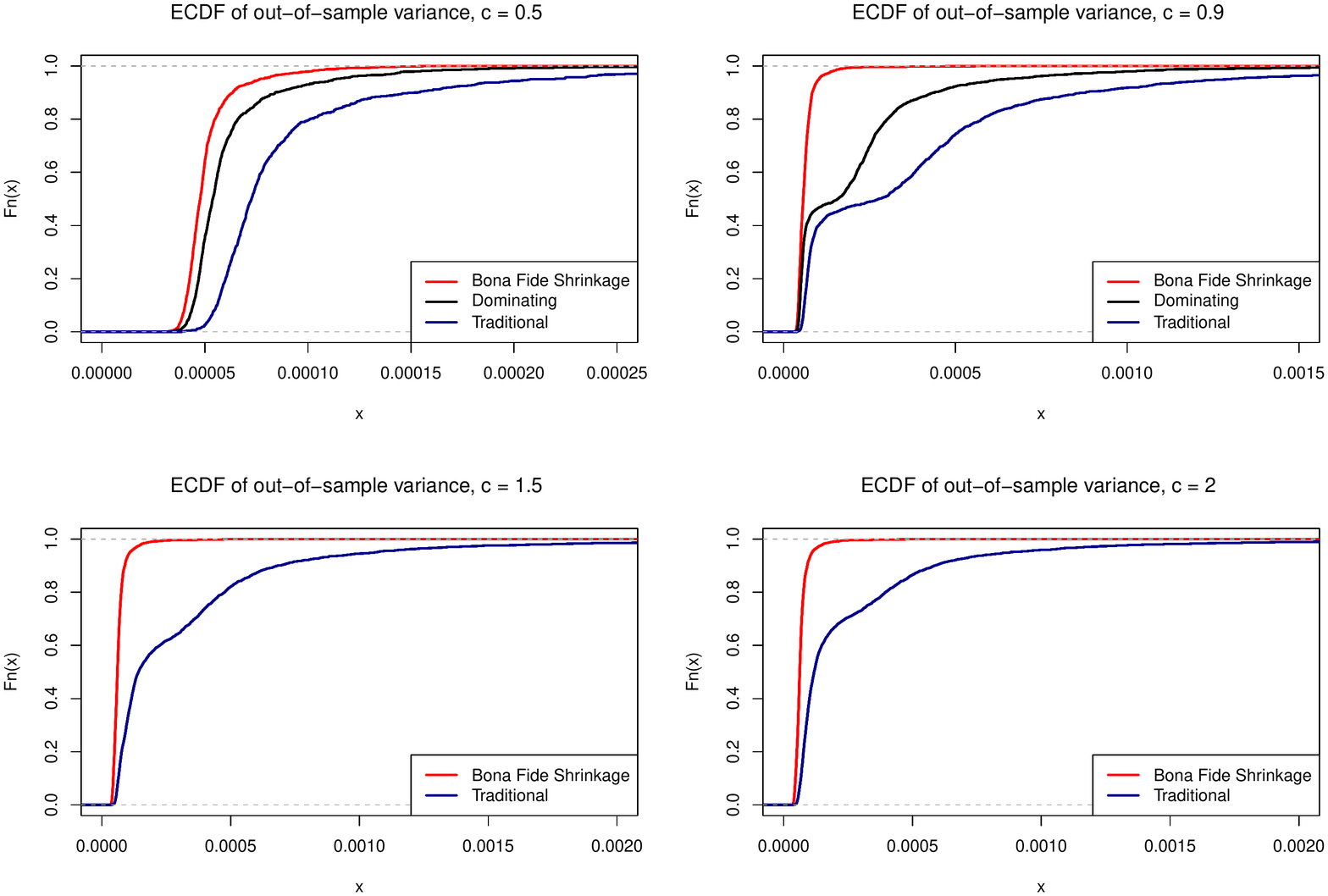}}
\vspace{-15mm}
\caption{Empirical distribution function of the out-of-sample variance for the bona fide optimal shrinkage estimator for the GMV portfolio together with the dominating and the traditional estimators.}
\label{Fig:ECDF_var}
\end{figure}
\end{landscape}

\begin{landscape}
\begin{figure}[h!tb]
\center\scalebox{0.8}{
\includegraphics[]{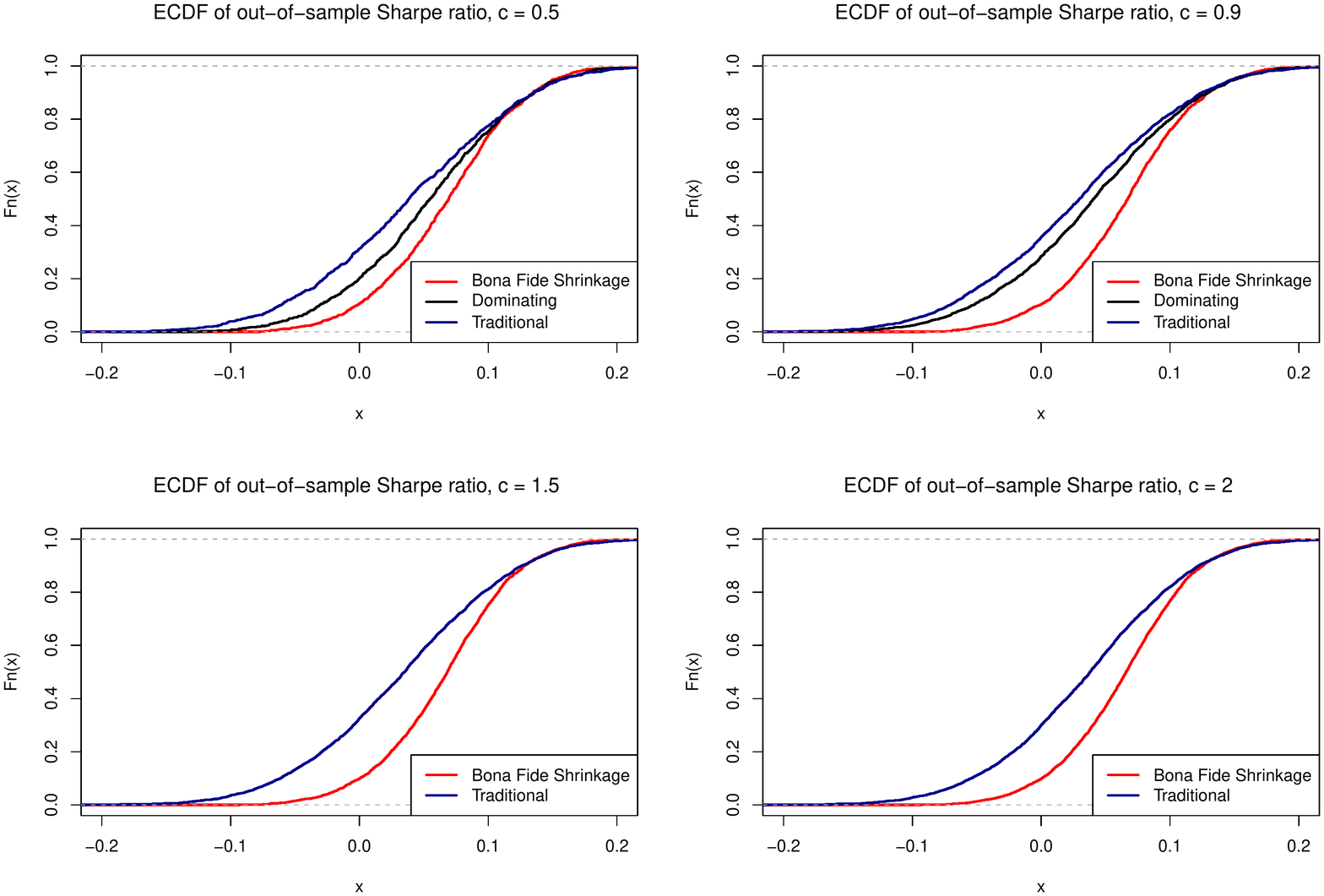}}
\vspace{-15mm}
\caption{Empirical distribution function of the out-of-sample Sharpe ratio for the bona fide optimal shrinkage estimator for the GMV portfolio together with the dominating and the traditional estimators.}
\label{Fig:ECDF_SR}
\end{figure}
\end{landscape}

\begin{center} Figures \ref{Fig:ECDF_var} and \ref{Fig:ECDF_SR} above here \end{center}

In Figures \ref{Fig:ECDF_var} and \ref{Fig:ECDF_SR} the e.c.d. functions of the out-of-sample variance and of the out-of-sample Sharpe ratio are presented for three estimators of the GMV portfolio, namely for the optimal shrinkage estimator, the traditional estimator and the dominating estimator suggested by Frahm and Memmel (2010). Because, the dominating estimator can be constructed only in case $c<1$, we drop it for $c=1.5$ and $c=2$. In all of the considered cases, we observe a very good performance of the optimal shrinkage estimator. It overperforms the other estimation strategies for both considered criteria. The corresponding e.c.d.f. lies above the other e.c.d. functions in case of the out-of-control variance, whereas it is below the e.c.d. functions of other competitors in case of the out-of-sample Sharpe ratio. On the second place, we rank the dominating estimator of Frahm and Memmel (2010) which is always better than the traditional estimator.

\section{Summary}

The global minimum variance portfolio plays an important role in investment theory and practice. This portfolio is widely used as an investment opportunity in both static and dynamic optimal portfolio choice problems. Although an explicit analytical expression for the structure of the GMV portfolio weights is available in literature, the estimation of the GMV portfolio appears to be a very challenging problem, especially for high-dimensional data.

We deal with this problem in the present paper by deriving a feasible and robust estimator for the weights of the GMV portfolio when the distribution of the asset returns is not prespecified and no market structure is imposed. We construct an optimal shrinkage estimator for the GMV portfolio which is optimal in the sense of minimizing the out-of-sample variance. An analytical expression for the shrinkage intensity is obtained which appears to be a complicated function of the data and the parameters of the asset return distribution. We deal with the later problem by determining an asymptotically equivalent quantity of the shrinkage intensity under high-dimensional asymptotics. We estimate this asymptotically equivalent function consistently by applying recent results from random matrix theory. This is achieved under very weak assumptions imposed on the distribution of the asset returns. Namely, we only require the existence of the fourth moment, whereas no explicit distributional assumption is imposed. Moreover, our findings are still valid in both cases $c<1$ and $c>1$ as well as if the spectrum of the population covariance matrix is bounded or unbounded. As a result, the suggested method can be applied to heavy-tailed distributed asset returns as well as to asset returns whose dynamics can be modeled by a factor model which is a very popular approach in financial and econometric literature. Finally, using simulated and real data, we compare the optimal shrinkage estimator for the GMV portfolio with existing ones. The theoretical findings as well as the results of the Monte Carlo simulations and an empirical study show that the suggested estimator for the GMV portfolio weights dominates the existing estimators in case $c>0$.

\section{Appendix} Here the proofs of the theorems are given. First, we point out that for our purposes $\bS_n$ can be well approximated by
\begin{equation*}\label{samplecov}
 \bS_n=\dfrac{1}{n}\bSigma_n^{\frac{1}{2}}\bx_n\left(\bI-\frac{1}{n}\bi\bi^\prime\right)\bx_n^{\prime}\bSigma_n^{\frac{1}{2}}
 \approx \dfrac{1}{n}\bSigma_n^{\frac{1}{2}}\bx_n\bx_n^{\prime}\bSigma_n^{\frac{1}{2}}\,,
\end{equation*}
since the matrix $\dfrac{1}{n^2}\bSigma_n^{\frac{1}{2}}\bx_n\bi\bi^\prime\bx_n^{\prime}\bSigma_n^{\frac{1}{2}}$ has rank one and, consequently, it does not influence the asymptotic behavior of the spectrum of the sample covariance matrix (see, Bai and Silverstein (2010), Theorem A.44).

Next, we present an important lemma which is a special case of Theorem 1 in Rubio and Mestre (2011).

\vspace{0.5cm}
\begin{lemma}\label{lem1}
Assume (A1) and (A2). Let a nonrandom $p\times p$-dimensional matrix $\mathbf{\Theta}_p$ possess a uniformly bounded trace norm (sum of singular values) and let $\bSigma_n=\bI$. Then it holds that
\begin{equation}\label{RM2011_id}
\left|\text{tr}\left(\mathbf{\Theta}_p(\bS_n-z\bI_p)^{-1}\right)-(x(z)-z)^{-1}\text{tr}\left(\mathbf{\Theta}_p\right)\right|\stackrel{a.s.}{\longrightarrow}0~~\text{for}~ p/n\longrightarrow c \in (0, +\infty) ~ \text{as} ~n\rightarrow\infty\,,
\end{equation}
where
\begin{equation}\label{RM2011_id_xz}
x(z)=\dfrac{1}{2}\left(1-c+z+\sqrt{(1-c+z)^2-4z}\right)\,.
\end{equation}
\end{lemma}

\noindent\textbf{Proof of Lemma \ref{lem1}:}
The application of Theorem 1 in Rubio and Mestre (2011) leads to (\ref{RM2011_id}) where $x(z)$ is a unique solution in $\mathbbm{C}^+$ of the following equation
\begin{equation}\label{eq1-Lemma6_1}
\dfrac{1-x(z)}{x(z)}=\dfrac{c}{x(z)-z}\,.
\end{equation}
The two solutions of (\ref{eq1-Lemma6_1}) are given by
\begin{equation}\label{solx}
x_{1,2}(z)=\dfrac{1}{2}\left(1-c+z\pm\sqrt{(1-c+z)^2-4z}\right)\,.
\end{equation}
In order to decide which of the two solutions is feasible, we note that $x_{1,2}(z)$ is the Stieltjes transform with a positive imaginary part. Thus, without loss of generality, we can take $z=1+c+i2\sqrt{c}$ and get
{\small
\begin{equation}\label{im}
\textbf{Im}\{x_{1,2}(z)\}=\textbf{Im}\left\{\dfrac{1}{2}\left(2+i2\sqrt{c}\pm i2\sqrt{2c}\right)\right\}=\textbf{Im}\left\{1+i\sqrt{c}(1\pm\sqrt{2})\right\}=\sqrt{c}\left(1\pm\sqrt{2}\right)\,,
\end{equation}
}
which is positive only if the sign $"+"$ is chosen. Hence, the solution is given by
\begin{equation}\label{solx_a}
x(z)=\dfrac{1}{2}\left(1-c+z+\sqrt{(1-c+z)^2-4z}\right)\,.
\end{equation}
Lemma \ref{lem1} is proved.\\

Rubio and Mestre (2011) studied the asymptotics of the functionals $\text{tr}(\mathbf{\Theta}(\bS_n-z\bI)^{-1})$ for a deterministic matrix $\mathbf{\Theta}$ with bounded trace norm at infinity. Note that the results of Theorem 1 of Rubio and Mestre (2011) also hold under the weaker assumption of the existence of the $4$th moments.
This statement is obtained by using Lemma B.26 of Bai and Silverstein (2010) on quadratic forms which we cite for presentation purposes as Lemma \ref{lem2} below.

\begin{lemma}\label{lem2}\textbf{[Lemma B.26, Bai and Silverstein (2010)]}
 Let $\bA$ be a $p\times p$ nonrandom matrix and let $\bx=(x_1,\ldots,x_p)^\prime$ be a random vector with independent entries. Assume that $E(x_i)=0$, $E|x_i|^2=1$, and $E|x_i|^l\leq\nu_l$. Then, for any $k\geq1$,
 \begin{equation}\label{BSlem}
 E|\bx^\prime\bA\bx-\text{tr}(\bA)|^k\leq C_k\left((\nu_4\text{tr}(\bA\bA^\prime))^{\frac{k}{2}}+\nu_{2k}\text{tr}(\bA\bA^\prime)^{\frac{k}{2}}\right)\,,
 \end{equation}
 where $C_k$ is some constant which depends only on $k$.
 \end{lemma}

In order to obtain the statement of Theorem 1 of Rubio and Mestre (2011) under the weaker assumption imposed on the moments, we replace Lemma 2 of Rubio and Mestre (2011) by Lemma \ref{lem2} in the case of $k\geq1$. Note that Lemma 2 of Rubio and Mestre (2011) holds for $k>1$ while Lemma \ref{lem2} is a more stronger result because it holds also in the case $k=1$. This is the main trick which implies that Lemma 3 of Rubio and Mestre (2011) holds also for $k\geq1$ (instead of $k>1$). Lemma 4 of Rubio and Mestre (2011) has already been proved under the assumption that there exist $4+\varepsilon$ moments. The last step is the application of Lemma 1, 2 and 3 of Rubio and Mestre (2011) with $k\geq1$. Finally, it can be easily checked that the further steps of the proof of Theorem 1 of Rubio and Mestre (2011) hold under the existence of $4+\varepsilon$ moments. In order to save space we leave the detailed technical proof of this assertion to the reader.

\vspace{1cm}
\textbf{Proof of Theorem \ref{th1}:}
Let us recall the optimal shrinkage intensity expressed as
 \begin{eqnarray}\label{alfa_app}
  \alpha_n^*&=& \dfrac{\mathbf{b}_n^{\prime}\bSigma_n\mathbf{b}_n-\dfrac{\bi^\prime\bS_n^{-1}\bSigma_n\mathbf{b}_n}{\bi^\prime\bS_n^{-1}\bi}}
 {\dfrac{\bi^\prime\bS_n^{-1}\bSigma_n\bS_n^{-1}\bi}{(\bi^\prime\bS_n^{-1}\bi)^2}-2\dfrac{\bi^\prime\bS_n^{-1}\bSigma_n\mathbf{b}_n}
 {\bi^\prime\bS_n^{-1}\bi}+\mathbf{b}_n^{\prime}\bSigma_n\mathbf{b}_n}\,.
 \end{eqnarray}
It holds that
\begin{eqnarray}
\bi^\prime\bS_n^{-1}\bi&=&\lim\limits_{z\rightarrow0^+}\text{tr}\left[(\bS_n-z\bSigma_n)^{-1}\bi\bi^\prime\right]\nonumber\\
&=&\lim\limits_{z\rightarrow0^+}\text{tr}\left[\left(\dfrac{1}{n}\bx_n\bx^\prime_n-z\bI\right)^{-1}\bSigma^{-\frac{1}{2}}_n\bi\bi^\prime\bSigma^{-\frac{1}{2}}_n\right]
\label{th31_eq1}\\
\bi^\prime\bS_n^{-1}\bSigma_n\mathbf{b}_n&=&\lim\limits_{z\rightarrow0^+}\text{tr}\left[(\bS_n-z\bSigma_n)^{-1}\bSigma_n\mathbf{b}_n\bi^\prime\right]\nonumber\\
&=&\lim\limits_{z\rightarrow0^+}\text{tr}\left[\left(\dfrac{1}{n}\bx_n\bx^\prime_n-z\bI\right)^{-1}\bSigma^{\frac{1}{2}}_n \mathbf{b}_n\bi^\prime\bSigma^{-\frac{1}{2}}_n\right]
\label{th31_eq2}\\
\bi^\prime\bS_n^{-1}\bSigma_n\bS_n^{-1}\bi&=& \left.\dfrac{\partial}{\partial z}\text{tr}\left[\left(\frac{1}{n}\bS_n-z\bSigma_n\right)^{-1}\bi\bi^\prime\right]\right|_{z=0}\nonumber\\
&=&\left.\dfrac{\partial}{\partial z}\text{tr}\left[\left(\frac{1}{n}\bx_n\bx^\prime_n-z\bI\right)^{-1} \bSigma^{-\frac{1}{2}}_n\bi\bi^\prime\bSigma^{-\frac{1}{2}}_n\right]\right|_{z=0}
\label{th31_eq3}\,.
\end{eqnarray}

Let
\begin{equation*}
\xi_n(z)=\text{tr}\left[\left(\dfrac{1}{n}\bx_n\bx^\prime_n-z\bI\right)^{-1}\mathbf{\Theta}_{\xi}\right]
\quad \text{with} \quad \mathbf{\Theta}_{\xi}=\bSigma^{-\frac{1}{2}}_n\bi\bi^\prime\bSigma^{-\frac{1}{2}}_n
\end{equation*}
and
\begin{equation*}
\zeta_n(z)=\text{tr}\left[\left(\dfrac{1}{n}\bx_n\bx^\prime_n-z\bI\right)^{-1}\mathbf{\Theta}_{\zeta}\right]
\quad \text{with} \quad \mathbf{\Theta}_{\zeta}=\bSigma^{\frac{1}{2}}_n\mathbf{b}_n\bi^\prime\bSigma^{-\frac{1}{2}}_n\,.
\end{equation*}
Both the matrices $\mathbf{\Theta}_{\xi} $ and $\mathbf{\Theta}_{\zeta}$ possess a bounded trace norm since
\begin{equation*}
\|\mathbf{\Theta}_{\xi}\|_{tr}
=\bi^\prime \bSigma^{-1}_n\bi\le M_l^{-1}
\end{equation*}
and
\begin{equation*}
\|\mathbf{\Theta}_{\zeta}\|_{tr}
=\sqrt{\bi^\prime \bSigma^{-1}_n\bi}\sqrt{\mathbf{b}_n^\prime \bSigma_n \mathbf{b}_n}\le \sqrt{\dfrac{M_u}{M_l}} \,.
\end{equation*}

Then, for all $z\in\mathbbm{C}^+$,  we get from Lemma \ref{lem1}
\begin{equation}\label{th31_eq4}
|\xi_n(z)-(x(z)-z)^{-1}\text{tr}\left[\mathbf{\Theta}_{\xi}\right]|=|\xi_n(z)-(x(z)-z)^{-1}\bi^\prime \bSigma^{-1}_n\bi| \stackrel{a.s.}{\longrightarrow}0~~\text{for}~ p/n\rightarrow c>0~ \text{as} ~n\rightarrow\infty
\end{equation}
and
\begin{equation}\label{th31_eq5}
|\zeta_n(z)-(x(z)-z)^{-1}\text{tr}\left[\mathbf{\Theta}_{\zeta}\right]|=\left|\zeta_n(z)-(x(z)-z)^{-1}\right|\stackrel{a.s.}{\longrightarrow}0~~\text{for}~ p/n\rightarrow c>0~ \text{as} ~n\rightarrow\infty\,,
\end{equation}
where $x(z)$ is given in (\ref{RM2011_id_xz}). Using that $\lim\limits_{z\rightarrow0^+}(x(z)-z)^{-1}=(1-c)^{-1}$ and combining (\ref{th31_eq4}) and (\ref{th31_eq5}) with (\ref{th31_eq1}) and (\ref{th31_eq2}) leads to
\begin{equation}\label{th31_eq6}
|\bi^\prime\bS_n^{-1}\bi-(1-c)^{-1} \bi^\prime \bSigma^{-1}_n\bi| \stackrel{a.s.}{\longrightarrow}0~~\text{for}~ p/n\rightarrow c>0~ \text{as} ~n\rightarrow\infty
\end{equation}
and
\begin{equation}\label{th31_eq7}
\left|\bi^\prime\bS_n^{-1}\bSigma_n\mathbf{b}_n-(1-c)^{-1}\right|\stackrel{a.s.}{\longrightarrow}0~~\text{for}~ p/n\rightarrow c>0~ \text{as} ~n\rightarrow\infty\,.
\end{equation}

Finally, using the equality
\begin{equation}\label{derxx}
\left.\dfrac{\partial}{\partial z}\dfrac{1}{ x(z)-z}\right|_{z=0}=-\left.\dfrac{x^\prime(z)-1}{ (x(z)-z)^2}\right|_{z=0}
=-\left.\dfrac{\dfrac{1}{2}\left(1-\frac{1+c-z}{\sqrt{(1-c+z)^2-4z}}\right)-1}{ (x(z)-z)^2}\right|_{z=0}
= \dfrac{1}{(1-c)^3}\,,
\end{equation}
we get
\begin{equation}\label{th31_eq8}
\left|\xi_n^\prime(0)-\left.\dfrac{\partial}{\partial z}(x(z)-z)^{-1}\right|_{z=0}\text{tr}\left[\mathbf{\Theta}_{\xi}\right]\right|=|\xi_n(z)-(1-c)^{-3}\bi^\prime \bSigma^{-1}_n\bi| \stackrel{a.s.}{\longrightarrow}0
\end{equation}
for $p/n\rightarrow c>0$ as $n\rightarrow\infty$. Consequently,
\begin{equation}\label{th31_eq9}
|\bi^\prime\bS_n^{-1}\bSigma_n\bS_n^{-1}\bi-(1-c)^{-3}\bi^\prime \bSigma^{-1}_n\bi| \stackrel{a.s.}{\longrightarrow}0~~\text{for}~ p/n\rightarrow c>0~ \text{as} ~n\rightarrow\infty\,.
\end{equation}

The application of (\ref{th31_eq6}) and (\ref{th31_eq9}) leads to
\begin{equation*}
\sigma_S^2 \stackrel{a.s.}{\longrightarrow} \frac{(1-c)^{-3}\bi^\prime \bSigma^{-1}_n\bi}{(1-c)^{-2}(\bi^\prime \bSigma^{-1}_n\bi)^2}=(1-c)^{-1}\sigma_{GMV}^2 ~~\text{for}~ p/n\rightarrow c>0~ \text{as} ~n\rightarrow\infty\,,
\end{equation*}
whereas additionally using (\ref{th31_eq7}) we get $\alpha_n^* \stackrel{a.s.}{\longrightarrow} \alpha^*$ with
\begin{equation*}
\alpha^*=\dfrac{\mathbf{b}_n^{\prime}\bSigma_n\mathbf{b}_n-\dfrac{(1-c)^{-1}}{(1-c)^{-1}\bi^\prime\bSigma_n^{-1}\bi}}
 {(1-c)^{-1}\sigma_{GMV}^2-2\dfrac{(1-c)^{-1}} {(1-c)^{-1}\bi^\prime\bSigma_n^{-1}\bi}+\mathbf{b}_n^{\prime}\bSigma_n\mathbf{b}_n}
= \dfrac{(1-c)R_{\mathbf{b}}}{c+(1-c)R_{\mathbf{b}}}
\end{equation*}
for $p/n\rightarrow c>0$ as $n\rightarrow\infty$. The quantity $R_{\mathbf{b}}$ is the limit of $R_{\mathbf{b}_n}$ which exists due to assumption $\sigma^2_{\mathbf{b}_n}\leq M_u$ and $\sigma^2_{GMV}\geq M_l$. These two equalities complete the proof of Theorem \ref{th1}.

\vspace{1cm}

\noindent\textbf{Proof of Theorem \ref{th12}:} In case of $c>1$, the optimal shrinkage intensity is given by
 \begin{eqnarray}\label{alfa_app_th12}
  \alpha_n^+&=& \dfrac{\mathbf{b}_n^{\prime}\bSigma_n\mathbf{b}_n-\dfrac{\bi^\prime\bS_n^{*}\bSigma_n\mathbf{b}_n}{\bi^\prime\bS_n^{*}\bi}}
 {\dfrac{\bi^\prime\bS_n^{*}\bSigma_n\bS_n^{*}\bi}{(\bi^\prime\bS_n^{*}\bi)^2}-2\dfrac{\bi^\prime\bS_n^{*}\bSigma_n\mathbf{b}_n}
 {\bi^\prime\bS_n^{*}\bi}+\mathbf{b}_n^{\prime}\bSigma_n\mathbf{b}_n}\,.
 \end{eqnarray}
Let $\mathbf{\Theta}_{\xi}=\bSigma^{-\frac{1}{2}}_n\bi\bi^\prime\bSigma^{-\frac{1}{2}}_n$ and $\mathbf{\Theta}_{\zeta}=\bSigma^{\frac{1}{2}}_n\mathbf{b}_n\bi^\prime\bSigma^{-\frac{1}{2}}_n$. Using the definition of $\bS_n^*$ given in \eqref{geninverse} and the equality $(\bx_n\bx_n^\prime)^+=\bx_n\left(\bx_n^\prime\bx_n\right)^{-2}\bx_n^\prime$, we get
\begin{eqnarray*}
\bi^\prime\bS_n^{*}\bi&=&\text{tr}\left[\left(\bx_n\bx_n^\prime\right)^+\mathbf{\Theta}_{\xi}\right]
=\text{tr}\left[\bx_n\left(\bx_n^\prime\bx_n\right)^{-2}\bx_n^\prime\mathbf{\Theta}_{\xi}\right] \nonumber\\
&=&\left.\dfrac{\partial}{\partial z}\text{tr}\left[\bx_n\left(\bx_n^\prime\bx_n-z\bI_n\right)^{-1}\bx_n^\prime\mathbf{\Theta}_{\xi}\right]\right|_{z=0}\\
\bi^\prime\bS_n^{*}\bSigma_n\mathbf{b}_n&=&\text{tr}\left[\left(\bx_n\bx_n^\prime\right)^+\mathbf{\Theta}_{\zeta}\right]
=\text{tr}\left[\bx_n\left(\bx_n^\prime\bx_n\right)^{-2}\bx_n^\prime\mathbf{\Theta}_{\zeta}\right]\nonumber\\
&=&\left.\dfrac{\partial}{\partial z}\text{tr}\left[\bx_n\left(\bx_n^\prime\bx_n-z\bI_n\right)^{-1}\bx_n^\prime\mathbf{\Theta}_{\zeta}\right]\right|_{z=0}\\
\bi^\prime\bS_n^{*}\bSigma_n\bS_n^{*}\bi&=& \text{tr}\left[\left(\left(\bx_n\bx_n^\prime\right)^+\right)^{-2}\mathbf{\Theta}_{\zeta}\right]
=\text{tr}\left[\bx_n\left(\bx_n^\prime\bx_n\right)^{-3}\bx_n^\prime\mathbf{\Theta}_{\xi}\right]\nonumber\\
&=&\frac{1}{2}\left.\dfrac{\partial^2}{\partial z^2}\text{tr}\left[\bx_n\left(\bx_n^\prime\bx_n-z\bI_n\right)^{-1}\bx_n^\prime\mathbf{\Theta}_{\xi}\right]\right|_{z=0}\,.
\end{eqnarray*}

The application of the Woodbury formula (matrix inversion lemma, see, e.g., Horn and Johnson (1985))
\begin{eqnarray}\label{wf}
\bx_n\left(\bx_n^\prime\bx_n-z\bI_n\right)^{-1}\bx_n^\prime=\bI_p+z\left(\bx_n\bx_n^\prime-z\bI_p\right)^{-1}
\end{eqnarray}
leads to
\begin{eqnarray*}
\bi^\prime\bS_n^{*}\bi&=&\left.\dfrac{\partial}{\partial z}z\text{tr}\left[\left(\bx_n\bx_n^\prime-z\bI_p\right)^{-1}\mathbf{\Theta}_{\xi}\right]\right|_{z=0}\\
\bi^\prime\bS_n^{*}\bSigma_n\mathbf{b}_n&=&\left.\dfrac{\partial}{\partial z} z \text{tr}\left[\left(\bx_n\bx_n^\prime-z\bI_p\right)^{-1}\mathbf{\Theta}_{\zeta}\right]\right|_{z=0}\\
\bi^\prime\bS_n^{*}\bSigma_n\bS_n^{*}\bi&=& \frac{1}{2}\left.\dfrac{\partial^2}{\partial z^2}z\text{tr}\left[\left(\bx_n\bx_n^\prime-z\bI_p\right)^{-1}\mathbf{\Theta}_{\xi}\right]\right|_{z=0}\,.
\end{eqnarray*}

From the proof of Theorem \ref{th1} we know that both the matrices $\mathbf{\Theta}_{\xi} $ and $\mathbf{\Theta}_{\zeta}$ possess the bounded trace norm. Then the application of Lemma \ref{lem1} leads to
\begin{eqnarray}
\bi^\prime\bS_n^{*}\bi&\stackrel{a.s.}{\longrightarrow}&\left.\dfrac{\partial}{\partial z}\frac{z}{x(z)-z}\right|_{z=0}\text{tr}\left[\mathbf{\Theta}_{\xi}\right]=
\left.\dfrac{\partial}{\partial z}\frac{z}{x(z)-z}\right|_{z=0}\bi^\prime \bSigma^{-1}_n\bi
\label{th12_eq1}\\
\bi^\prime\bS_n^{*}\bSigma_n\mathbf{b}_n&\stackrel{a.s.}{\longrightarrow}&\left.\dfrac{\partial}{\partial z}\frac{z}{x(z)-z}\right|_{z=0}\text{tr}\left[\mathbf{\Theta}_{\zeta}\right]=\left.\dfrac{\partial}{\partial z}\frac{z}{x(z)-z}\right|_{z=0}\label{th12_eq2}\\
\bi^\prime\bS_n^{*}\bSigma_n\bS_n^{*}\bi&\stackrel{a.s.}{\longrightarrow}& \frac{1}{2}\left.\dfrac{\partial^2}{\partial z^2}\frac{z}{x(z)-z}\right|_{z=0}\text{tr}\left[\mathbf{\Theta}_{\xi}\right]=\frac{1}{2}\left.\dfrac{\partial^2}{\partial z^2}\frac{z}{x(z)-z}\right|_{z=0}\bi^\prime \bSigma^{-1}_n\bi\label{th12_eq3}
\end{eqnarray}
for $p/n\rightarrow c>1$ as $n\rightarrow\infty$, where $x(z)$ is given in \eqref{RM2011_id_xz}.

Let us make the following notations
\begin{equation}\label{not}
\theta(z)=\dfrac{z}{x(z)-z} \quad \text{and} \quad \phi(z)=\dfrac{x(z)-zx^\prime(z)}{z^2}\,.
\end{equation}
Then the first and the second derivatives of $\theta(z)$ are given by
\begin{equation}\label{not_der}
\theta^{\prime}(z)=\theta^2(z)\phi(z)\quad \text{and} \quad \theta^{''}(z)=2\theta(z)\theta^\prime(z)\phi(z)+\theta^2(z)\phi^\prime(z)\,.
\end{equation}

Using L'Hopital's rule, we get
\begin{equation}\label{tetalim}
\theta(0)=\lim\limits_{z\rightarrow0^+}\theta(z)=\lim\limits_{z\rightarrow0^+}\dfrac{z}{x(z)-z}=\lim\limits_{z\rightarrow0^+}\dfrac{1}{(x^\prime(z)-1)}=\dfrac{1}{\dfrac{1}{2}\left(1-\dfrac{1+c}{|1-c|}\right)-1}=-\dfrac{c-1}{c}\,,
\end{equation}
\begin{equation}\label{philim}
\phi(0)=\lim\limits_{z\rightarrow0^+}\phi(z)=\lim\limits_{z\rightarrow0^+}\dfrac{x(z)-zx^\prime(z)}{z^2}=-\dfrac{1}{2}\lim\limits_{z\rightarrow0^+}x^{''}(z)=-\dfrac{1}{2}\lim\limits_{z\rightarrow0^+}\frac{-2c}{((1-c+z)^2-4z)^{3/2}}
=\dfrac{c}{(c-1)^3}\,,
\end{equation}
\begin{eqnarray*}
\lim\limits_{z\rightarrow0^+}\phi^\prime(z)&=&-\lim\limits_{z\rightarrow0^+}\dfrac{2(x(z)-zx^\prime(z))+z^2x^{''}(z)}{z^2}\nonumber\\
&=&-\lim\limits_{z\rightarrow0^+}\dfrac{2\phi(z)+x^{''}(z)}{z}=-\lim\limits_{z\rightarrow0^+}(2\phi^\prime(z)+x^{'''}(z))\,,
\end{eqnarray*}
which implies
\begin{equation}\label{phi1lim2}
\phi^\prime(0)=\lim\limits_{z\rightarrow0^+}\phi^\prime(z)=-\dfrac{1}{3}\lim\limits_{z\rightarrow0^+}x^{'''}(z))==-\dfrac{1}{3}\lim\limits_{z\rightarrow0^+}\frac{6c(z-c-1)}{((1-c+z)^2-4z)^{5/2}}=\dfrac{2c(c+1)}{(c-1)^5}
\,.
\end{equation}

Combining (\ref{not_der}), (\ref{tetalim}), (\ref{philim}), and (\ref{phi1lim2}), we get
\begin{equation}\label{num11a}
\theta^{'}(0)=\lim\limits_{z\rightarrow0^+}\theta^{'}(0)=\theta^2(0)\phi(0)=\dfrac{1}{c(c-1)}\,,
\end{equation}
\begin{equation}\label{num11}
\theta^{''}(0)=\lim\limits_{z\rightarrow0^+}\theta^{''}(z)=\theta^3(0)\phi^2(0)+\theta^2(0)\phi^\prime(0)=\dfrac{2}{(c-1)^3}\,.
\end{equation}

Finally, the application of the last two equalities together with \eqref{th12_eq1}, \eqref{th12_eq2}, and \eqref{th12_eq3}, leads to
\begin{equation*}\label{ainfty+}
\alpha^+_n\overset{a.s.}{\longrightarrow}\alpha^+~~\text{for} ~~ \dfrac{p}{n}\rightarrow c\in(1, +\infty)~~ \text{as} ~~ n\rightarrow\infty\,,
\end{equation*}
where
\[
\alpha^+= \dfrac{\mathbf{b}_n^{\prime}\bSigma_n\mathbf{b}_n-\dfrac{\theta^{'}(0)}{\theta^{'}(0)\bi^\prime \bSigma^{-1}_n\bi}}
 {\dfrac{\theta^{''}(0)\bi^\prime \bSigma^{-1}_n\bi}{(\theta^{'}(0)\bi^\prime \bSigma^{-1}_n\bi)^2}-2\dfrac{\theta^{'}(0)}{\theta^{'}(0)\bi^\prime \bSigma^{-1}_n\bi}+\mathbf{b}_n^{\prime}\bSigma_n\mathbf{b}_n}
 =\dfrac{(c-1)R_{\mathbf{b}}}{(c-1)^2+c+(c-1)R_{\mathbf{b}}}~~\text{with}~R_{\mathbf{b}_n}\to R_{\mathbf{b}}\,
\]
and
\begin{equation}\label{outsampvariance+}
\sigma^2_{S^*}\overset{a.s.}{\longrightarrow}\dfrac{\theta^{''}(0)\bi^\prime \bSigma^{-1}_n\bi}{(\theta^{'}(0)\bi^\prime \bSigma^{-1}_n\bi)^2}=\dfrac{c^2}{c-1}\sigma_{GMV}^2 ~~\text{for} ~~ \dfrac{p}{n}\rightarrow c\in(1, +\infty)~~ \text{as} ~~ n\rightarrow\infty\,.
\end{equation}
This completes the proof of Theorem \ref{th12}.\\

\noindent\textbf{Proof of Theorem \ref{th2}}.
First of all,  we note that the asymptotic distribution of the quantity $\bi^\prime\bS_n^{-1}\bi$ has already been derived in Theorem \ref{th1}. From (\ref{th31_eq6}) we get that a consistent estimator of $\bi^\prime\bSigma_n^{-1}\bi$ is given by
\begin{eqnarray}\label{con}
&&\widehat{\bi^\prime\bSigma_n^{-1}\bi}=(1-p/n)\bi^\prime\bS_n^{-1}\bi~~\text{for}~c<1,\\
&&\widehat{\bi^\prime\bSigma_n^{-1}\bi}=p/n(p/n-1)\bi^\prime\bS_n^{*}\bi~~\text{for}~c>1\,.
\end{eqnarray}
In order to complete the proof of Theorem \ref{th2} we need the following lemma of Rubio and Mestre (2011) \\[0.5cm]

\begin{lemma}\label{lem3}
Let $\{\bol{\xi}_1,\ldots,\bol{\xi}_n\}$ be a sequence of i.i.d. real random vectors with zero mean and unit variance and with uniformly bounded $4+\varepsilon$ moments for some $\varepsilon>0$ and let $\bC_n$ be some nonrandom matrix with bounded trace norm at infinity. Then it holds that
\begin{equation}\label{lemma}
\left|\dfrac{1}{n}\sum\limits_{i=1}^n\bol{\xi}_i^\prime\bC_n\bol{\xi}_i-\text{tr}(\bC_n) \right|\stackrel{a.s.}{\longrightarrow}0~~\text{for}~ p/n\longrightarrow c \in (0, +\infty) ~ \text{as} ~n\rightarrow\infty \,.
\end{equation}
\end{lemma}

\vspace{0.5cm}
Next, we rewrite $\mathbf{b}_n^\prime\bS_n\mathbf{b}_n$ in the following way
\begin{equation}\label{rew}
\mathbf{b}_n^\prime\bS_n\mathbf{b}_n=\dfrac{1}{n}\sum\limits_{i=1}^n\bol{x}_i^\prime\bSigma_n^{\frac{1}{2}}\mathbf{b}_n\mathbf{b}_n^\prime\bSigma_n^{\frac{1}{2}}\bol{x}_i=\dfrac{1}{n}\sum\limits_{i=1}^n\bol{x}_i^\prime\bR_n\bol{x}_i\,
\end{equation}
where $\bol{x}_i$ is the $i$th column of the matrix $\bx_n$. For the application of Lemma \ref{lem3} we have to show that the matrix $\bR_n$ has a bounded trace norm at infinity. It holds that
\begin{equation}\label{Rnorm}
||\bR_n||_{tr}
= \mathbf{b}_n^\prime \bSigma_n \mathbf{b}_n \le M_u
\end{equation}
and, hence, the boundedness of the trace norm of the matrix $\bR_n$ follows directly from the assumption $\mathbf{b}_n^\prime \bSigma_n \mathbf{b}_n \le M_u$. The application of Lemma \ref{lem3} leads to
\begin{equation}\label{rew}
\left|\mathbf{b}_n^\prime\bS_n\mathbf{b}_n-\mathbf{b}_n^\prime\bSigma_n\mathbf{b}_n \right|\stackrel{a.s.}{\longrightarrow}0~~\text{for}~ p/n\longrightarrow c \in (0,1) ~ \text{as} ~n\rightarrow\infty\,,
\end{equation}
which together with (\ref{con}) implies the statement of Theorem \ref{th2}.

\newpage

\vspace{-10mm}

\newpage

\vspace{-10mm}

\newpage

\newpage
\clearpage

\newpage
\clearpage

\end{document}